\documentclass[prd,twocolumn, nofootinbib,superscriptaddress,preprintnumbers]{revtex4}

\usepackage{placeins}
\usepackage{soul}
\usepackage{chngcntr}
\usepackage{color}
\usepackage{epsfig}  
\usepackage{graphicx}
\usepackage{tabularx}
\usepackage{xspace}
\usepackage{float}
\usepackage[usenames, dvipsnames]{xcolor}

\usepackage{color}
\usepackage{hyperref}
\hypersetup{
     colorlinks   = true,
     citecolor    = MidnightBlue,
	 linkcolor=MidnightBlue
}
\usepackage{halloweenmath}
\usepackage{verbatim}
\usepackage{amsmath}
\usepackage{amssymb}
\usepackage{mathtools}
\usepackage{url}
\usepackage{bbold}
\usepackage{slashed}
\usepackage{array}
\usepackage{comment}

\usepackage{multirow}
\usepackage{threeparttable}
\usepackage{paralist}
\usepackage{xspace}
\usepackage{upgreek}
\usepackage{lipsum}
\usepackage{mathrsfs}

\newcommand{\abs}[1]{\left| #1 \right|} 
\newcommand\colvec[3][]{\begin{pmatrix}\ifx\relax#1\relax\else#1\\\fi#2\\#3\end{pmatrix}}

\definecolor{darkmagenta}{rgb}{0.55, 0.0, 0.55}
\newcommand{\beq}{\begin{equation}}
\newcommand{\beqn}{\begin{eqnarray}}
\newcommand{\eeq}{\end{equation}}
\newcommand{\eeqn}{\end{eqnarray}}
\newcommand\numberthis{\addtocounter{equation}{1}\tag{\theequation}}

\newcommand\order[1]{{\cal O}#1}

\DeclareRobustCommand{\Eq}[1]{Eq.~(\ref{#1})}

\DeclareRobustCommand{\Sec}[1]{Sec.~\ref{#1}}
\DeclareRobustCommand{\App}[1]{Appendix~\ref{#1}}
\DeclareRobustCommand{\Fig}[1]{Fig.~\ref{#1}}

\DeclareMathAlphabet\mathbfcal{OMS}{cmsy}{b}{n}

\newcommand{\eV}{ \textrm{eV} }
\newcommand{\keV}{ \textrm{keV} }

\newcommand{\cm}{ \textrm{cm} }
\newcommand{\km}{ \textrm{km} }

\newcommand{\Qv}{\mathcal{Q}_v}

\newcommand{\p}{\prime}
\newcommand{\x}{\text{\tiny MCP}}
\newcommand{\xb}{\overline{\text{\tiny MCP}}}
\newcommand{\w}{\omega}
\newcommand{\g}{\gamma}
\newcommand{\M}{\mathcal{M}}
\newcommand{\kvec}{\mathbf{k}}

\newcommand{\xv}{{\bf x}}

\newcommand{\vv}{{\bf v}}
\newcommand{\Ev}{{\bf E}}

\newcommand{\nl}{\nonumber \\}
\newcommand{\Ap}{A^\prime}
\newcommand{\mAp}{m_{A^\prime}}
\newcommand{\eps}{\epsilon}
\newcommand{\grad}{\nabla}
\newcommand{\para}{\mathbin{\!/\mkern-5mu/\!}}
\newcommand{\ind}{k}

\begin{document}
\preprint{MIT-CTP/5358}
\title{A Helioscope for Gravitationally Bound Millicharged Particles}
\author{Asher Berlin}
\email{ajb643@nyu.edu}
\affiliation{Center for Cosmology and Particle Physics, Department of Physics, New York University, New York, NY 10003, USA}
\affiliation{Theoretical Physics Department, Fermilab, P.O. Box 500, Batavia, IL 60510, USA}
\author{Katelin Schutz}\thanks{Einstein Fellow}
\email{katelin.schutz@mcgill.ca}
\affiliation{Center for Theoretical Physics, Massachusetts Institute of Technology, Cambridge, MA 02139, USA}
\affiliation{Department of Physics \& McGill Space Institute, McGill University, Montr\'eal, QC H3A 2T8, Canada}

\begin{abstract}\noindent

\noindent Particles may be emitted efficiently from the solar interior if they are sufficiently light and weakly coupled to the solar plasma. In a narrow region of phase space, they are emitted with velocities smaller than the escape velocity of the solar system, thereby populating a gravitationally bound density that can accumulate over the solar lifetime, referred to as a ``solar basin." Detection strategies that can succeed in spite of (or even be enhanced by) the low particle velocities are therefore poised to explore new regions of parameter space when taking this solar population into account. Here we identify ``direct deflection" as a powerful method to detect such a population of millicharged particles. This approach involves distorting the local flow of gravitationally bound millicharges with an oscillating electromagnetic field and measuring these distortions with a resonant LC circuit. Since it is easier to distort the flow of slowly moving particles, the signal is parametrically enhanced by the small solar escape velocity near Earth. The proposed setup can probe couplings an order of magnitude smaller than other methods for millicharge masses ranging from 100 meV to 100 eV and can operate concurrently as a search for sub-GeV millicharged dark matter. The signal power scales as the millicharge coupling to the eighth power, meaning that even with conservative assumptions, direct deflection could begin to explore new regions of parameter space. We also highlight novel features of millicharge solar basins, including those associated with the phase space distribution and the possibility for the occupation number to vastly exceed that of a thermal distribution. 
\end{abstract}

\maketitle

\section{Introduction}
\label{sec:intro}

Stellar interiors are an excellent probe of physics beyond the Standard Model (SM). Owing to their high density, temperature, and volume, the interaction rate inside of stars is extremely large, providing ample opportunities to produce weakly-coupled particles through rare processes. Moreover, if those particles have a sufficiently weak coupling to the SM plasma, they will be able to stream through the star and escape, an effect which has been extensively studied in the context of stellar energy loss~\cite{Raffelt:1996wa}. In analyzing the abundances and inferred lifetimes of various stellar populations, one can bound the stellar energy loss rate and consequently place extremely strong limits on the emission of, e.g., light axions, hidden photons and $B-L$ vectors, scalars and pseudoscalars coupled to nucleons and electrons, and millicharged\footnote{Here we do not refer to particles with charge of order $10^{-3} e$ but rather follow the naming convention for referring to particles with charges $q_\x \ll 1$. We note that other nonmenclature is sometimes used to refer to the same particles, e.g., ``minicharged.''} particles (MCPs). Stellar energy loss bounds provide some of the strongest constraints on sub-keV particles in extensions of the SM, and with a few exceptions (such as dark photons~\cite{An:2013yfc, Redondo:2013lna, Hardy:2016kme} and sterile neutrinos~\cite{Raffelt:2011nc, Arguelles:2016uwb}) these bounds are strong down to arbitrarily low particle masses.

\begin{figure*}[t]
\includegraphics[width=\textwidth]{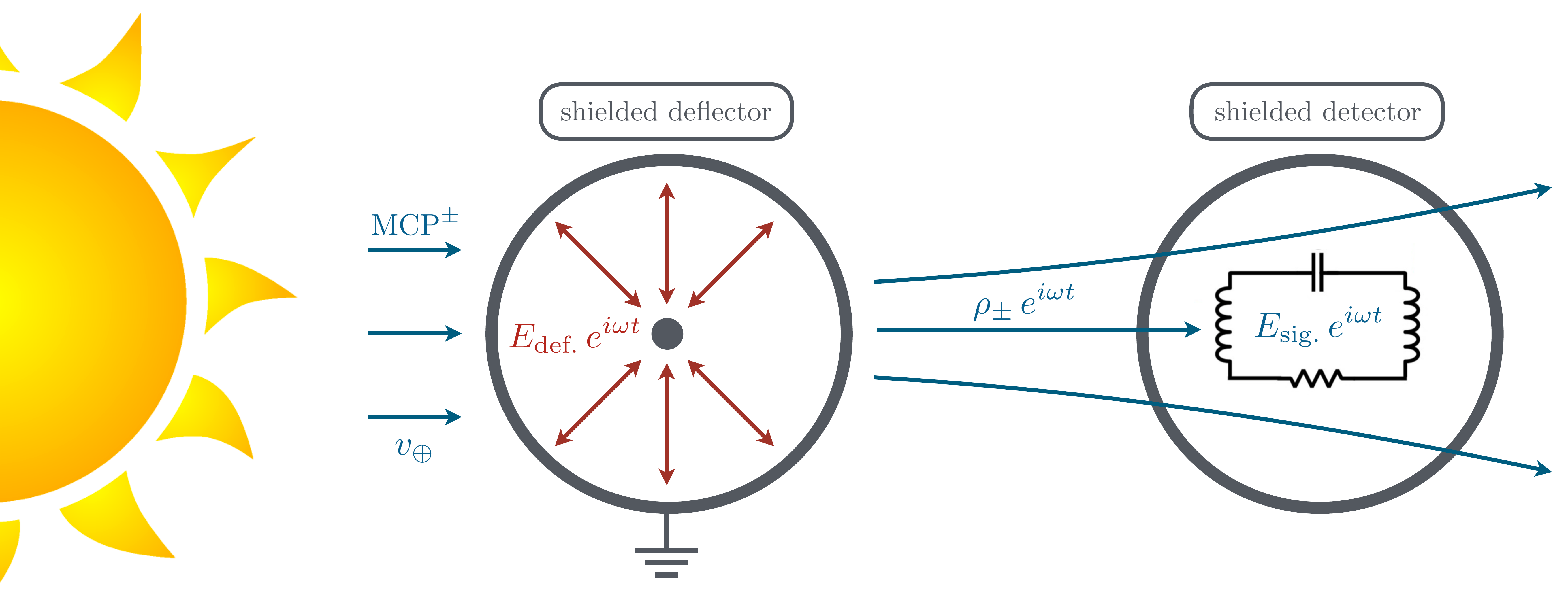}
\vspace{-0.6cm}
\caption{A schematic of the ``direct deflection" helioscope setup. A gravitationally bound population of MCPs is created from plasmon decay in the solar interior, building up a dense ``solar basin" within the solar system over gigayear timescales. In the Earth frame, a ``wind" of these MCPs is generated by the relative orbital motion of the Earth at a velocity of $v_\oplus \sim 10^{-4}$. In the terrestrial lab, the local charge-symmetric millicharge density passes into a shielded ``deflector" region, in which a driven electric field $E_\text{def.}$ oscillating with frequency $\omega$ induces a propagating wave train of oscillating millicharge overdensities $\rho_\pm$. Downwind, these charge densities penetrate a quiet shielded ``detector" region and source an oscillating electric field signal $E_\text{sig.}$ that is resonantly detected using an LC circuit tuned to the same frequency.}
\label{fig:cartoon}
\end{figure*}

In addition to considering stellar energy loss and its impact on stellar lifetimes, it may also be possible to directly detect the particles ejected from stars in a laboratory-based experiment. Due to its proximity to Earth, the Sun is an ideal stellar source of these particles, rendering such experiments ``helioscopes.'' Previous helioscopes have searched for a solar flux of sub-keV relativistic particles with energies comparable to the solar temperature, $T_\odot \sim 1 \ \keV$. However, as was recently shown in Refs.~\cite{VanTilburg:2020jvl,Lasenby:2020goo}, there may additionally be a large density of highly \emph{non-relativistic} particles produced from the Sun and bound gravitationally to the solar system, forming a ``solar basin.'' Typically, for processes occurring at the keV energy scale, the emission of a non-relativistic particle with mass $m \ll1\,$keV is highly phase-space suppressed. However, the accumulation of gravitationally bound particles over billions of years can compensate for such phase space suppression. In fact, for certain particle masses, the gravitationally bound density  \emph{exceeds} the predicted density of relativistic particles.

One obstacle to the detection of these non-relativistic particles is that their kinetic energy at Earth is at most $\sim 10\ \mu \eV \times (m/\keV)$, since the local escape velocity of the solar system is $\sim 10^{-4}$. Such low energies are well below the threshold of existing and proposed detectors~\cite{Lin:2019uvt}, indicating that single-particle elastic scattering processes involving solar basin particles are undetectable. However, in certain theories the limited kinetic energy of these particles may be circumvented by using their rest mass as a way to exceed the energy threshold of a detector, enabling sensitivity down to $\eV$-scale masses~\cite{VanTilburg:2020jvl,Lasenby:2020goo}. For instance, inelastic single-particle processes, such as absorption, are detectable so long as the mass gap is above threshold, meaning that some kinds of basin particles can be detected using standard dark matter direct detection targets such as xenon-based experiments~\cite{Akerib:2017uem,Fu:2017lfc,aprile2019light,aprile2020observation}, DAMIC~\cite{Aguilar-Arevalo:2016zop}, and CDMS~\cite{Bloch:2016sjj}.
However, such absorption processes are forbidden if gauge or spacetime symmetries forbid particle number violating interactions, such as in the simplest theories of MCPs. In this case, different detection strategies are needed, especially those which may be able to explore new parameter space in spite of or even \emph{because of} the low velocity of gravitationally bound particles.

\begin{figure*}[t]
\includegraphics[width= 1.3 \columnwidth]{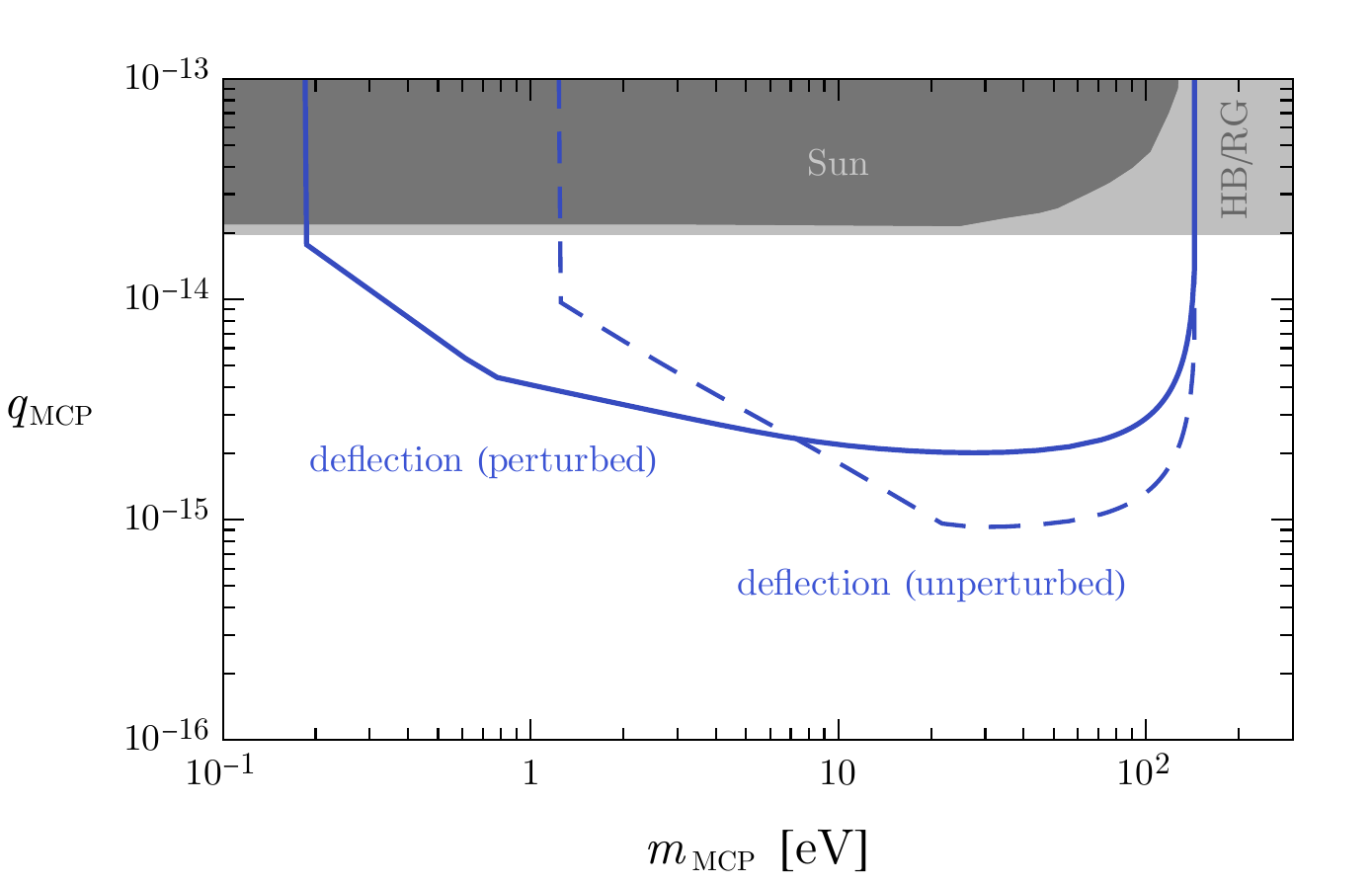}
\caption{The projected sensitivity (blue lines) of a direct deflection setup to a solar basin of fermionic MCPs compared to existing constraints from stellar energy loss (shaded gray)~\cite{Davidson:2000hf,Vinyoles:2015khy}. In each case, we assume an experimental integration time $t_\text{int.} = 1 \ \text{yr}$, a deflector/detector volume $V_\text{def.} = V_\text{det.} = 10 \ \text{m}^3$, a deflector electric field $E_\text{def.} = 10 \ \text{kV} / \cm$ oscillated at an angular frequency of $\w = 10 \ \text{kHz}$, and an LC circuit detector with quality factor $Q_\text{LC} = 10^7$ operating at a temperature $T_\text{LC} = 10 \ \text{mK}$. The solid and dashed blue lines assume a basin phase space that is maximally or minimally perturbed by gravitational encounters, respectively. We restrict our sensitivity projections to regions of parameter space in which the density is greater than $1/(10 \ \cm^3)$, so that the solar basin of MCPs can be treated as a continuum. For a basin density that does not saturate the upper bound from \Sec{sec:saturation} and assuming a deflector and detector of comparable size, the sensitivity scales with the experimental parameters as $q_\x \propto V_\text{def.}^{-7/24} \, E_\text{def.}^{-1/4} \, T_\text{LC}^{1/8} \, (\w \, t_\text{int.} \, Q_\text{LC})^{-1/8}$.}
\label{fig:reach}
\end{figure*}

In this article, we focus on a solar basin of MCPs in the $\eV-\keV$ mass range and identify ``direct deflection'' as the optimal strategy to directly detect this solar basin. This strategy (which was recently introduced in Ref.~\cite{Berlin:2019uco} in the context of dark matter detection) involves inducing and subsequently detecting oscillating overdensities of MCPs using large driven fields and well-shielded precision resonant detectors. A schematic of this approach is shown in Fig.~\ref{fig:cartoon}. Applied to MCPs in the solar basin, ambient MCPs passing through a shielded region containing a driven oscillating electric field are deflected, setting up propagating wave trains of MCP charge density. These charge overdensities penetrate a downstream electromagnetic shield, establishing a small oscillating electric field that can be measured with a resonant detector. A unique qualitative feature of this technique is that it probes the collective effects of the large number density of particles, instead of relying on the energy deposition from a single-particle scattering event. Relatedly, since more slowly moving particles are more easily deflected by the driven electric field, the signal is parametrically enhanced by the reduced kinetic energy of the solar basin compared to the typical kinetic energy of virialized dark matter. Thus, a direct deflection helioscope is extremely well-suited for the detection of a MCP solar basin. The most stringent existing constraints on sub-keV MCPs are derived from considerations of stellar energy loss, which exclude MCP charges larger than $q_\x \simeq 2 \times 10^{-14}$~\cite{Davidson:2000hf,Vinyoles:2015khy}. Our projected sensitivity of a direct deflection setup -- which extends beyond existing constraints for $100 \ \text{meV} - 100 \ \eV$ MCP masses -- is shown in Fig.~\ref{fig:reach}. This projected sensitivity also extends well beyond the reach of conventional helioscope searches for the elastic scattering of the relativistic flux of solar MCPs, whose current  sensitivity is significantly weaker than existing astrophysical limits~\cite{Khan:2020vaf}.

The rest of this article is organized as follows. We review the parameter space of MCPs in \Sec{sec:summary} and present a conceptual overview of directly deflecting the MCP solar basin (including order-of-magnitude scaling arguments) in order to provide intuition for our main results. In \Sec{sec:phasespaceproduction}, we show a first calculation of the full phase space distribution of particles in the solar basin. In \Sec{sec:saturation} we  point out that if production of the solar basin proceeds through emission of multiple dark sector particles per interaction (as is the case for MCPs), then the occupation number of bound particles can greatly exceed that of a distribution that is in equilibrium with the Sun. In the case of fermions, the density saturates because of Pauli blocking, whereas the occupation number of bosons can greatly exceed unity. We leave the detailed study of constraining millicharged bosons (and other bosons with multiple particles produced per interaction) to future work~\cite{KVT}, primarily focusing on constraining fermionic MCPs in this article. We outline the main solar production mechanisms for MCPs in \Sec{sec:solarproduction}. In \Sec{sec:deflection}, we highlight the experimental approach of direct deflection as a means to discover the MCP basin and illustrate the projected sensitivity of such a setup. In \Sec{sec:interactions}, we demonstrate how MCP interactions with the solar environment or amongst themselves may alter the characteristics of the solar basin. Finally, in \Sec{sec:conclusion}, we conclude and discuss directions for future inquiry. A series of appendices is also provided that contains additional details of our calculations. In \App{app:MCPproduction}, we calculate the solar production rate of MCPs, \App{app:chargedensity} outlines some details associated to the calculation of the direct deflection signal, and in \App{app:hydro}, we point out that if MCP self-interactions efficiently drive the basin towards a state of hydrostatic equilibrium, then the density profile of the solar basin can be significantly altered. 

\section{Model Space and Conceptual Overview}
\label{sec:summary}

MCPs possess a small effective electromagnetic charge $q_\x \ll 1$ and naturally arise in models where  a light dark photon $\Ap$ kinetically mixes with SM electromagnetism. In this case, particles charged directly under the dark photon appear as millicharged under normal electromagnetism on length-scales smaller than the dark photon Compton wavelength with an effective charge  $q_\x = \eps \, e^\p / e$, where $\eps \ll 1$ is the kinetic mixing parameter, $e^\p$ is the dark photon gauge coupling, and $e$ is the SM electric charge~\cite{Holdom:1985ag}. If $\eps$ is generated radiatively from loops of $N^\p$ generations of heavy particles charged under both the SM and dark sector, the natural expectation for the strength of the kinetic mixing is $\eps \sim N^\p \, e^\p e / (4 \pi)^2$ (although certain models predict parametrically smaller values~\cite{Gherghetta:2019coi}). Considerations of stellar energy loss exclude MCP couplings larger than $q_\x \simeq 2 \times 10^{-14}$ for masses $m_\x \lesssim 1 \ \keV$~\cite{Davidson:2000hf,Vinyoles:2015khy}. Hence, for values of $\eps$ near the  radiative estimate above, the dark sector fine structure constant $\alpha^\p \equiv e^{\p \, 2} / 4 \pi \sim 4 \pi q_\x / N^\p$ needs to be smaller than $\sim 10^{-13} / N^\p$, which controls the strength of MCP self-interactions. Although very small gauge couplings appear unnatural in a top-down framework of  gauge coupling unification~\cite{Gherghetta:2019coi}, they are theoretically consistent and can arise in LARGE volume string compactifications~\cite{Burgess:2008ri}. Therefore, it behooves us to consider small MCP self-couplings. As we will show, this implies that the MCP stellar basin is long-lived and can survive to the present time. However, self-interactions can still be important. In fact, we show in \App{app:hydro} that for sufficiently large $\alpha^\p$, self-scatters can modify the local density and spatial profile of the solar basin. 

Independent of MCPs, direct constraints on light dark photon mediators (such as those derived from stellar energy loss~\cite{An:2013yfc,Redondo:2013lna}, laboratory tests of Coulomb's law~\cite{Williams:1971ms}, and measurements of the cosmic microwave background~\cite{Mirizzi:2009iz,Caputo:2020bdy}) decouple as the dark photon mass is taken to zero, thus motivating the consideration of ultralight or massless dark photons. In this work, we focus on dark photon masses $\mAp \lesssim 10^{-8} \ \eV \sim (10 \ \text{m})^{-1}$, in which case the MCPs effectively couple to standard electromagnetism over macroscopic length-scales~\cite{Berlin:2019uco}. From the point of view of solar production and terrestrial detection, we thus treat such particles as electromagnetically charged. 

MCPs that are produced in the solar interior will only remain gravitationally bound in a narrow region of phase space. The density of emitted MCPs that satisfy this criterion is significantly peaked near the Sun, due to its large gravitational attraction. As a result, the terrestrial density of the MCP solar basin is suppressed both by the large Earth-to-Sun distance as well as the small region of phase space below the solar escape velocity. As we show in detail in \Sec{sec:solarproduction}, the dominant production mechanism for MCPs is through the decay of electromagnetic plasmon excitations in the solar interior. For this process, the terrestrial number density $n (r_\oplus)$ of gravitationally bound MCPs is largest for masses $m_\x$ comparable to the solar plasma frequency $\w_p \sim 100 \ \eV$, such that 
\begin{align}
\label{eq:approxdensity}
n(r_\oplus) &\sim \Big[ \frac{\alpha_\text{em} \, q_\x^2 \, \w_p^4}{4 \pi^3} ~ \frac{r_\odot^3 \, t_\odot}{r_\oplus^3} \Big]  \times \Big[v_\text{esc.} (r_\odot)  \, v_\text{esc.}(r_\oplus)^2 \Big] \, 
\nl
&\sim 10^5 \ \cm^{-3} \times \left( \frac{q_\x}{2 \times 10^{-14}} \right)^2
~,
\end{align}
where $r_\odot \simeq 7 \times 10^5 \ \km \simeq 5 \times 10^{-3} \ \text{AU}$ is the solar radius,  $r_\oplus \simeq 1 \ \text{AU}$ is the Earth's distance from the Sun, $v_\text{esc.} (r)$ is the solar escape velocity at heliocentric radius $r$, and $t_\odot \simeq 4.5 \times 10^9 \ \text{yr}$ is the age of the solar system. In the first line of \Eq{eq:approxdensity}, the first set of brackets is the local density of MCPs produced over a solar lifetime, assuming that all such MCPs remain gravitationally bound. In particular, the ratio $r_\odot^3 t_\odot/r_\oplus^3$ arises from assuming that the entire volume of the Sun can produce particles over the whole lifetime of the Sun and that those particles get redistributed to the volume within 1~AU of the Sun; the factor of $\omega_p^4$ then arises to give dimensions of number density (recall that here we have chosen $m_\x\sim \w_p$, making $\w_p$ the only dimensionful quantity that determines the particle production rate inside the Sun). The second set of brackets accounts for the fact that these particles only get bound to within 1~AU of the Sun for a small kinematic subset of emitted MCPs, corresponding to the fraction of phase space with MCP velocity smaller than $v_\text{esc.}(r_\odot)$ yet sufficiently large to make it to Earth, i.e.,
\beq
 v_\text{esc.}(r_\odot) - \frac{v_\text{esc.}(r_\oplus)^2}{2 \, v_\text{esc.}(r_\odot)}\lesssim v_\x \lesssim v_\text{esc.}(r_\odot)
~.
\eeq
We can think of the ``velocity volume'' of this three-dimensional kinematic phase space as occupying a thin spherical shell of radius $v_\text{esc.}(r_\odot)$ and thickness $v_\text{esc.}(r_\oplus)^2/2 \, v_\text{esc.}(r_\odot)$ over which the production rate does not vary, giving rise to the scaling in the second set of brackets of the first line of \Eq{eq:approxdensity}. In the second line of \Eq{eq:approxdensity}, we have fixed the MCP coupling to saturate existing constraints from stellar energy loss, as discussed above. 

Direct deflection is an especially powerful detection strategy for particles with low velocities, which tends to enhance the overall strength of the signal~\cite{Berlin:2019uco}. From \Eq{eq:approxdensity}, we see that even for very small couplings, the density of these particles is quite large, such that we can describe the MCP solar basin as a continuum (i.e., using continuous variables like the mean density and ignoring Poisson fluctuations in the local number of particles). In the solar frame, there is no bulk ``flow" (as distinct from the motions of individual particles) of the MCP basin. However, the relative motion of the Earth's orbit leads to a headwind of MCPs flowing in the opposite direction in Earth's frame, analogous to the dark matter wind from the galactic motion of our solar system. A simplified schematic of the terrestrial laboratory setup is shown in \Fig{fig:cartoon}, which consists of two regions (a ``deflector" and ``detector") surrounded by electromagnetic shields. Inside the deflector region, a large electric field $E_\text{def.}$ is driven at frequency $\w \lesssim 10 \ \text{kHz}$. As the ``wind" of MCPs flows unimpeded into this region (due to the small coupling), the electric field induces a wave train of small MCP charge densities $\rho_\pm$ oscillating at the same frequency that propagate into a quiet downwind detection region. Inside the detector, these MCP charge densities source a small oscillating signal electric field $E_\text{sig.}$ that can be resonantly detected with an LC circuit tuned to the same frequency. 

In \Sec{sec:deflection}, we provide a technical description of the induced charge densities and resulting signal. Here, we give a brief summary of the parametrics to provide an intuitive picture of the signal. The most straightforward way to derive the form of the MCP charge density $\rho_\pm$ is to first consider the electric force on an individual MCP as it traverses the interior of the deflector. Provided that $\w$ is sufficiently small, this MCP sees an effectively static electric field. 
For simplicity, if we imagine the initial MCP velocity in Earth's frame to be the orbital velocity of Earth $v_\oplus$ and the initial trajectory of this particle to be perfectly aligned with the deflector-detector axis, then this MCP gets a perpendicular ``kick" from $E_\text{def.}$, such that this new component to its velocity is $v_\pm \sim \pm e q_\x \, (E_\text{def.} / m_\x) \, (R_\text{def.} / v_\oplus)$, 
where $\pm$ corresponds to the sign of the MCP's charge and $R_\text{def.}$ is the characteristic length-scale of the deflector. Since the MCP basin is charge symmetric, MCPs of either sign contribute to a net current density oscillating at $\w$ with amplitude
\begin{align}
j_\pm &\sim e q_\x \, n (r_\oplus) \, (v_+ - v_-) 
\nl
&\sim (e q_\x)^2 \, (n (r_\oplus) / m_\x) \, (\varphi_\text{def.} / v_\oplus)
~,
\end{align}
where $\varphi_\text{def.} \sim E_\text{def.} \, R_\text{def.}$ is the electric potential of the deflector, and we have assumed that $v_\pm / v_\oplus \ll 1$ such that the number density is approximately unperturbed. By charge continuity, such a current density implies a corresponding charge density of amplitude
\begin{align}
\label{eq:approxcharge}
\rho_\pm &\sim - j_\pm / v_\oplus \sim - \frac{(e q_\x)^2 \, n (r_\oplus)}{m_\x \, v_\oplus^2} ~ \varphi_\text{def.} 
\nl
&\sim - m_{D, \x}^2 ~ \varphi_\text{def.}
~,
\end{align}
where $m_{D, \x}$ is the MCP contribution to the photon's Debye mass. The last equality in the expression above is the standard result for how a weakly-coupled plasma (the MCP basin) Debye screens a quasi-static electric source (the deflector)~\cite{Lifshitz:99987}. However, unlike standard Debye screening, in this case these charge densities exist even in regions where the deflector electric potential vanishes, e.g., inside the detector shield placed downwind. This is because the MCP charge densities that develop in the non-zero electric potential of the deflector region are swept outside of the deflector by the MCP wind. A key feature of \Eq{eq:approxcharge} is that unlike traditional scattering-based detection experiments, the signal in a direct deflection setup does not fall below experimental thresholds at small kinetic energies; in fact, it is enhanced at small velocities.\footnote{Note, though, that the small solar escape velocity at Earth $v_\text{esc.} (r_\oplus) \sim v_\oplus$ suppresses the solar production rate, i.e., $n (r_\oplus) \propto v_\text{esc.} (r_\oplus)^2$ as in \Eq{eq:approxdensity}, such that the MCP charge density $\rho_\pm \propto n (r_\oplus) / v_\oplus^2$ is approximately independent of $v_\oplus$ or $v_\text{esc.}(r_\oplus)$. However, the fact that the signal does not fall off as $v_\text{esc.}(r_\oplus) \to 0$ is a unique advantage of a direct deflection setup.}

These MCP charge densities source a real oscillating electric field $E_\text{sig.} \sim \rho_\pm \, R_\text{def.}$ of size
\begin{align}
\label{eq:approxEsig}
E_\text{sig.} &\sim 10^{-17} \ \text{kV} \ \cm^{-1} \times \left( \frac{q_\x}{2 \times 10^{-14}} \right)^{2} \left( \frac{m_\x}{100 \ \eV} \right)^{-1}
\nl
&\times \left( \frac{n (r_\oplus)}{10^5 \ \cm^{-3}} \right)  \left( \frac{R_\text{def.}}{1 \ \text{m}} \right) \left( \frac{\varphi_\text{def.}}{1 \ \text{MV}} \right) 
~,
\end{align}
inside the detection region, which drives a signal current in an LC circuit resonantly tuned to the same frequency. In \Eq{eq:approxEsig}, we have normalized the MCP model parameters to be consistent with \Eq{eq:approxdensity}. As we discuss in \Sec{sec:deflection}, a meter-sized cryogenic LC circuit optimized to detect such electric fields can measure oscillating fields as small as $\sim 10^{-21} \ \text{kV} \ \cm^{-1}$, thereby enabling impressive sensitivity to currently unexplored parameter space. In the remainder of this paper, we provide a detailed derivation of the basin density and direct deflection signal, which is ultimately needed to derive the projected sensitivity shown in \Fig{fig:reach}.

\begin{figure*}[t]
\centering
\includegraphics[width=\textwidth]{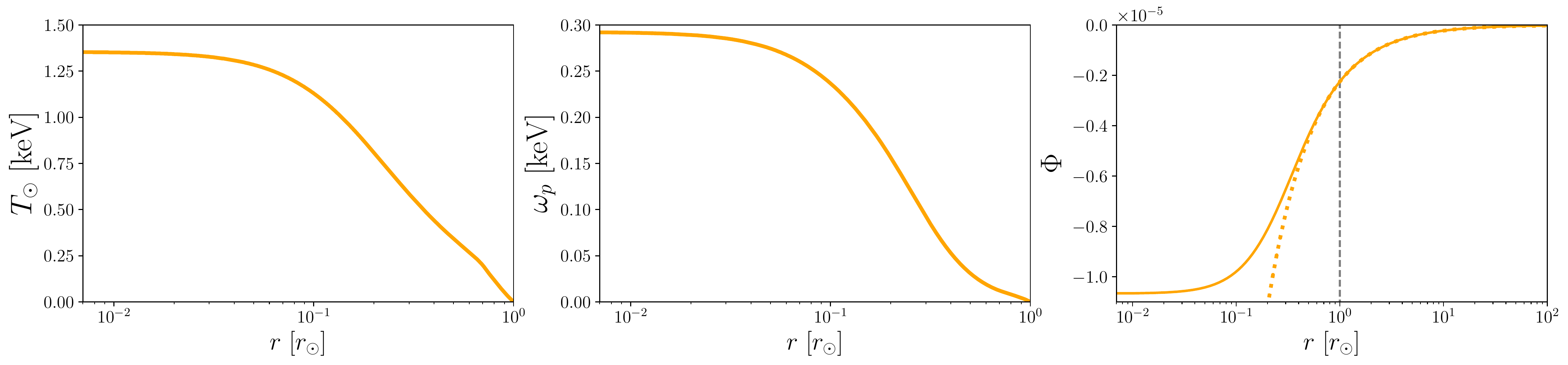}
\caption{The temperature (left), plasma frequency (middle), and gravitational potential (right; note the scale of $10^{-5}$) radial profiles of the Sun. The gravitational potential matches onto the $1/r$ scaling, shown as a dotted line, outside the radius of the Sun, marked by a vertical dashed line.}
\label{fig:Sunstuff}
\end{figure*}

\section{Phase Space Density of a Gravitationally Bound Population}
\label{sec:basin}

A derivation of the total integrated number density of particles in the solar basin was given in Ref.~\cite{VanTilburg:2020jvl}. In \Sec{sec:phasespaceproduction}, we provide an alternative derivation, which for the first time allows for the extraction of the raw basin phase space density from stellar production, which is of particular importance for understanding the physics of detecting a basin of MCPs. In particular, the strength of MCP self-interactions, which can potentially alter the basin density, as well as the detailed nature of the direct deflection signal, are both intimately tied to the structure of the MCP phase space. 

Later in \Sec{sec:saturation}, we point out a feature of the phase space evolution that is unique to basin particles that are pair-produced, like MCPs. In particular, for the models considered in Refs.~\cite{VanTilburg:2020jvl,Lasenby:2020goo}, the phase space saturates at thermal occupancies. However, for models with higher multiplicity production of basin particles (e.g., multiple MCPs in the final state), the density instead saturates near the degenerate limit for fermionic basin particles and possibly at very high occupation numbers for theories involving light bosons. Additional dynamics of the solar basin, such as those stemming from self-interactions, are discussed in detail towards the end of this work (see \Sec{sec:interactions}). 

\subsection{Phase Space from Production}
\label{sec:phasespaceproduction}

In this subsection, we present a derivation (that differs from Ref.~\cite{VanTilburg:2020jvl}) that allows for the extraction of the basin phase space distribution as a function of heliocentric radius $r$. The results for the velocity distribution $f$ in \Eq{eq:f1} and the number density $n$ in Eqs.~(\ref{eq:nexact1}) and (\ref{eq:napprox1}) are the main results of this subsection.

The gravitationally bound MCP population is emitted from the Sun with a speed less than the solar escape velocity, $v_\text{esc.} (r_\odot) = \order(10^{-3})$ depending on the exact location inside the Sun. Therefore, we work in the non-relativistic limit where the luminosity of emitted MCPs per solar volume, $Q = dL /d^3 \xv^\p$, is an isotropic quantity that can be expressed as\footnote{In this section, we adopt the notation in which primed quantities denote properties of the MCP at production; e.g., $t^\p$, $\xv^\p$, and $\vv^\p$ are the time, position, and velocity of the MCP upon production in the solar interior, with the spatial origin being located at the center of the Sun. Unprimed variables pertain to properties of the MCP after it has escaped the solar interior; e.g., $r$ will be used to denote the radial distance of the MCP from the center of the Sun after production. Later, starting in \Sec{sec:deflection}, we use vectors with tildes, such as $\tilde{\xv}$ and $\tilde{\xv}^\p$, to denote position vectors with respect to the experimental apparatus, i.e., where the spatial origin is located at the center of the deflector in \Fig{fig:cartoon}.}  
\beq
\label{eq:Qvdef}
\frac{dQ}{d^3 \vv^\p} = v^{\p \, 2( \ind -1)} ~ \Qv
~.
\eeq
Above, we have separated the dependence of the energy loss rate into a dependence on the particle's speed when it is produced, $v^\p$, with power-law index for integer $\ind \geq 1$, and a factor $\Qv$ that only depends on properties of the solar interior (e.g., the density and temperature) and on intrinsic parameters in the underlying particle physics model (e.g., the MCP coupling, mass, and  spin). We take the properties of the Sun to be spherically symmetric and constant in time so that $\Qv$ depends only on the radius $r^\p$ where the particle was produced. For the temperature and density of the Sun, we adopt the values from the Standard Solar Model of Ref.~\cite{Bahcall:2004fg}, with the radial profiles of some key quantities shown in \Fig{fig:Sunstuff}.

To derive the phase space density of particles at heliocentric radius $r$, we first compute the phase space density upon production in the solar interior using the definition of luminosity density in the non-relativistic limit, $Q \simeq m_\x \,  dN / d^3 \xv^\p \, dt^\p$, where $N$ is the total number of emitted MCPs. With this notation, the velocity phase space density $f(\xv, \vv, t^\p)$ at the time of production $t^\p$ is given by 
\begin{align*} 
\label{eq:dfprime}
d f(\xv, \vv, t^\p) = dN (\xv^\p, \vv^\p, t^\p) ~ \delta^3(\xv- \xv^\p) ~ \delta^3(\vv- \vv^\p)
& \\ = \frac{d^3 \xv^\p \, d^3 \vv^\p \, dt^\p}{m_\x} ~ \frac{d Q(\xv^\p, \vv^\p)}{d^3 \vv^\p}  ~ \delta^3(\xv- \xv^\p) ~ \delta^3(\vv- \vv^\p)
~ . 
\numberthis 
\end{align*}
To obtain the distribution at later times, we assume that the particles originating at position $\xv^\p$ free stream out of the solar interior and follow trajectories solely determined by the solar gravitational potential. For instance, we temporarily neglect effects such as additional gravitational interactions from planetary encounters in the solar system, reabsorption in the solar interior, or MCP self-interactions (these are discussed in detail in later sections). We thus time-evolve the primed arguments of the delta functions in \Eq{eq:dfprime} as
\begin{align*}
& \delta^3 \big( \xv- \xv') \rightarrow \delta^3 \big( \xv- \xv_\text{traj.}(\xv^\p, \vv^\p , t^\p; t) \big) \numberthis
\\& \delta^3(\vv- \vv') \rightarrow \delta^3 \big( \vv- \vv_\text{traj.}(\xv^\p , \vv^\p, t^\p; t) \big)
~,
\numberthis 
\end{align*}
where $\xv_\text{traj.}$ and $\vv_\text{traj.}$ are the time-evolved points in phase space corresponding to the orbital trajectory of a particle at time $t$ with initial conditions $\xv'$ and $\vv'$ at time $t' < t$. This time evolution is subject to a number of constraints given the symmetries of the problem. For instance, since the gravitational potential is taken to be spherically symmetric, conservation of the direction of the angular momentum vector implies that the trajectories are restricted to a plane that passes through the center of the Sun. For any such plane, we can therefore reduce the phase space dimensionality of the problem from six to four and sum together all such planes in order to integrate 
the total density. 

In any given plane, it is most convenient to work in polar coordinates $r$, $\theta$, $v_r = dr / dt$, and $v_\theta = d \theta / d t$, so that the time-evolved phase space can be expressed as 
\begin{align*} 
\label{eq:df1}
&df = dr^\p \, d\theta^\p \,dv_r^\p \, dv_\theta^\p \,dt^\p ~ \frac{v^{\p \,2(\ind-1)} \Qv}{m_\x} ~ \Big(\frac{r^\prime}{r_\text{traj.}}\Big)^2 \numberthis  
\\ &\times \delta(v_r - v_{r_\text{traj.}})\, \delta(v_\theta - v_{\theta_\text{traj.}}) \, \delta(\theta - \theta_\text{traj.}) \,  \delta(r - r_\text{traj.})
~,
\end{align*} 
where we have suppressed the arguments of the time-evolved trajectory variables. The first two of the four remaining delta functions in \Eq{eq:df1} can be reexpressed as constraints on the initial velocities $v_r^\p$ and $v_\theta^\p$ using conservation of energy
\beq
E = \frac{1}{2} \, m_\x \, v_{r_\text{traj.}}^2 + m_\x \, \Phi_\text{eff.}(r_\text{traj.})
\eeq
and conservation of the magnitude of the angular momentum
\beq
\ell = m_\x \, r_\text{traj.}^2 \, v_{\theta_\text{traj.}} 
\eeq
along the entire trajectory, where
\beq
\Phi_\text{eff.}(r) = \Phi(r) + \frac{\ell^2}{2 \, m_\x^2 \, r^2}
\eeq
 is the one-dimensional effective potential in the radial direction and $\Phi < 0$ is the solar gravitational potential. These constraints amount to making the following replacements in \Eq{eq:df1}
\begin{align}
\label{eq:vdeltas}
\delta(v_r - v_{r_\text{traj.}}) &= \frac{v_r}{v_{r 0}^\p} \, \big[ \delta(v_r^\p - v_{r 0}^\p ) + \delta(v_r^\p + v_{r 0}^\p ) \big]
\nl
\delta(v_\theta - v_{\theta_\text{traj.}}) &= \Big( \frac{r}{r^\p} \Big)^2 ~ \delta (v_\theta^\p - v_{\theta 0}^\p \, )
~,
\end{align}
where we have defined
\beq
\label{eq:vprime}
v_{r 0}^\p \equiv  \sqrt{2(E/m_\x - \Phi_\text{eff.}(r^\p))}
~~,~~ 
v_{\theta 0}^\p \equiv \Big(\frac{r}{r^\p}\Big)^2 ~ v_\theta
 ~.
\eeq
In the first line of \Eq{eq:vdeltas}, the delta functions enforce $v_{r_\text{traj.}}$ and $v_r$ to have the same sign in addition to the same magnitude; e.g., if $v_r > 0$, then the radial velocity delta function only has weight in regions of $t^\p$ for which the trajectory is outgoing at time $t$. 

 The angular delta function in \Eq{eq:df1} can be expressed as an orbit equation, obtained by solving for the integrals of motion $r$ and $\theta$ and eliminating the time coordinate,
\beq  
\delta(\theta - {\theta_\text{traj.}}) = \delta \Big(\theta^\p  - \theta + \frac{1}{m_\x} \, \int_{r^\p}^{r} dr^{\p \p} \frac{\ell}{r^{\p \p \, 2} \, v_{r^{\p\p}}} \Big)
~.
\eeq 
The final radial delta function in \Eq{eq:df1} can be used to eliminate the time integral. In rewriting this spatial delta function in terms of time $t^\p$, summing over all possible roots amounts to including all possible previous times $t^\p < t$ at which a particle was emitted at the right initial point in phase space to have evolved to the phase space point $(r, \theta, v_r, v_\theta)$ at time $t$. In particular, for gravitationally bound orbits ($E < 0$), we make the following replacement in \Eq{eq:df1}, 
\begin{align}
\label{eq:deltar}
\delta(r - r_\text{traj.}) &= \Theta(-E) \, \Theta(E-E_\text{min.}) \, \sum_{t_0^\p} \frac{\delta(t' - t_0^\p \, )}{\abs{v_r}} 
\nl
&= \Theta(-E) \, \Theta(E-E_\text{min.}) \, \frac{t}{2 \, t_\text{orb.}} ~ \frac{\delta(t')}{\abs{v_r}}
~,
\end{align}
where 
\beq
E_\text{min.} \equiv m_\x \, \max{(\Phi_\text{eff.}(r), \Phi_\text{eff.}(r^\p))}
~,
\eeq
and the sum is over all roots $t^\p = t_0^\p$ that satisfy $r = r_\text{traj.}(\xv^\p,  \vv^\p , t_0^\p \, ; t)$ and $v_r = v_{r_\text{traj.}} (\xv^\p,  \vv^\p , t_0^\p \, ; t)$. The step functions enforce that the orbit is gravitationally bound and also able to reach out to radius $r$. In the limit where the age of the solar basin $t$ is much longer than the orbital period, we have expressed the sum over the roots in \Eq{eq:deltar} as a product of a single delta function times the number of phase space crossings $t/(2 \, t_\text{orb.})$, where $t_\text{orb.}$ is the time it takes to get from the perihelion $r_\text{min.}$ to the aphelion $r_\text{max.}$ of the orbit,\footnote{Note that we have ignored an additional term in \Eq{eq:deltar} that is only relevant for unbound orbits, $E > 0$. In this case, the number of phase space crossings is instead just equal to unity, so that \Eq{eq:deltar} is replaced by $\delta(r - r_\text{traj.}) = \Theta(E) \delta(t') / \abs{v_r}$.} 
\beq 
\label{eq:torbit}
t_\text{orb.} = \int_{r_\text{min.}}^{r_\text{max.}} \frac{dr^{\p \p}}{\sqrt{2(E/m_\x - \Phi_\text{eff.}(r^{\p \p}))}}
~.
\eeq 
For general orbits, this must be calculated numerically since the gravitational potential inside the Sun does not scale as $1 / r$.

With the formalism outlined above, integrating \Eq{eq:df1} then yields the total velocity phase space at time $t$ for bound orbits of energy $E > E_\text{min.}$,
\beq
\label{eq:f1}
f (r, v_r , v_\theta) = \frac{t}{t_\text{orb.}} \int dr^\p  \frac{(v_0^\p)^{2 (\ind-1)}}{v_{r 0}^\p} \frac{\Qv (r^\p)}{m_\x}  \Theta \big(v_{r 0}^{\p \, 2} \big)
\, ,
\eeq
where we have defined $(v_0^\p )^2 \equiv (v_{r 0}^\p\, )^2 + (r^\p v_{\theta 0}^\p \, )^2$. Note that when integrating over $r^\p$, the solar production rate $\Qv$ only has weight for radii within the solar interior, $r ^\p< r_\odot$. 

 For any point $\xv$ outside the Sun, the planar phase space distribution of Eq.~\eqref{eq:f1} can be rotated azimuthally along the axis between the center of the Sun and $\xv$, due to cylindrical symmetry; thus, $|\xv| = r$ can be identified with the longitudinal cylindrical coordinate $z$ and $r \theta$ can be taken to be the transverse radial cylindrical coordinate $\rho$. The perihelion is constrained to be within the Sun where MCP production can occur, i.e., $r_\text{min.}<r_\odot$. This upper bound on $r_\text{min.}$ also bounds $v_\theta$ from above, since larger $v_\theta$ (and hence larger $\ell$) strengthens the corresponding centrifugal barrier in $\Phi_\text{eff.}$. As a result, the phase space density tends to be quite collimated along the radial direction in the limit that $r_\odot \ll r$. This can also be seen from the Heaviside step function $\Theta$ in the integrand of \Eq{eq:f1}, which enforces $(v_{r 0}^\p\, )^2 > 0$; from the definition of $v_{r 0}^\p$ in \Eq{eq:vprime}, this is equivalent to
\beq
\label{eq:vrhomax}
v_\rho^2 = (r \,  v_\theta)^2 <  (v_\rho^\text{max.})^2 \equiv  \frac{v_r^2 + 2 (\Phi(r) - \Phi(r^\p))}{(r/r^\p)^2 - 1}
~.
\eeq
Thus, far outside the Sun ($r \gg r^\p \sim r_\odot$), the transverse velocity is constrained to be $v_\rho \ll v_\text{esc.} (r_\odot)$, as expected. 

\begin{figure}[t]
\includegraphics[width=0.49\textwidth]{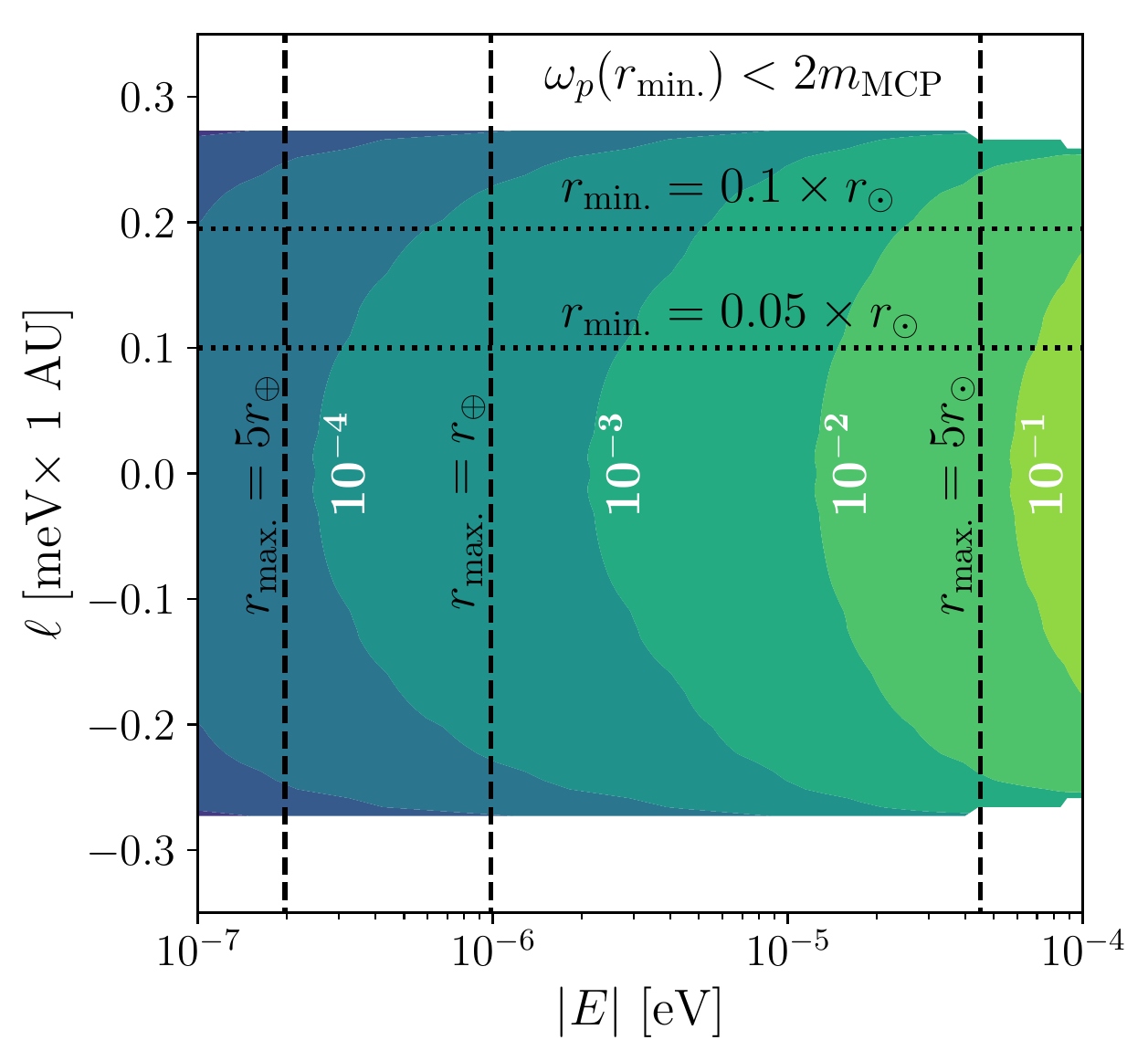}
\vspace{-0.6cm}
\caption{The momentum phase space density $f_p$ of a fermionic MCP solar basin for a mass and coupling of $m_\x = 100 \ \text{eV}$ and $q_\x = 10^{-16}$, as a function of orbital energy $E$ and angular momentum $\ell$. Note that $f_p$ is related to the velocity phase space $f$ by $f_p \simeq (2 \pi / m_\x)^3 \, (f/4)$ (see the discussion near \Eq{eq:fp}). The coupling $q_\x$ has been chosen to be sufficiently small such that $f_p \ll 1$ for the displayed values of $E$ and $\ell$. In this case, the phase space density scales as $f_p \propto q_\x^2$. For sufficiently large couplings, the density saturates at $f_p \simeq 1$ (see \Sec{sec:saturation}). Also shown as dotted  and dashed black contours are  values of the perihelion $r_\text{min.}$ and aphelion $r_\text{max.}$, respectively. For large values of $|\ell|$, the solar plasma mass is sufficiently small at radii larger than $r_\text{min.}$ such that plasmon decay to MCPs is kinematically forbidden.}
\label{fig:fp}
\end{figure}

In \Fig{fig:fp}, we show the phase space density from solar production as a function of energy $E$ and angular momentum $\ell$ for representative values of the MCP mass and coupling. Along the dotted and dashed lines we also show values of the perihelion $r_\text{min.}$ and aphelion $r_\text{max.}$, which are determined by $\ell$ and $E$, respectively. We find that the phase space density is enhanced for more radial and deeply bound orbits with aphelia closer to the Sun. We also note that as expected from Liouville's theorem, for a fixed choice of $E$ and $\ell$, the phase space density is independent of position $r$, provided that $r_\text{min.} \lesssim r \lesssim r_\text{max.}$ and energy and angular momentum are conserved.

The number density $n$ is obtained by integrating \Eq{eq:f1} over velocity. Switching variables from $v_r$ to $E$, this corresponds to
\begin{align}
\label{eq:nexact1}
& n (r, t) = \frac{2^\ind \, \pi \, t}{m_\x^2} \int dr^\p ~ \Qv (r^\p)  \int_0^{v_\rho^\text{max.}} dv_\rho ~ v_\rho
\nl
&  \int_{E_\text{min.}}^0 dE ~ \frac{ t_\text{orb.}^{-1} ~ \big[ \frac{E}{m_\x} - \Phi (r^\p)\big]^{\ind-1}}{\sqrt{\frac{E}{m_\x} - \Phi_\text{eff.} (r^\p)} \, \sqrt{\frac{E}{m_\x} - \Phi_\text{eff.} (r)}}
~,
\end{align}
where the energy integral is restricted to $E_\text{min.} \leq E \leq 0$. We can make further analytic progress by noting that very far away from the Sun the orbits are approximately radial, $E / m_\x \gg v_\rho^2$, in which case the aphelion is $r_\text{max.} \simeq - G \, M_\odot \, m_\x / E \gg r_\text{min.}$ and the orbital time of \Eq{eq:torbit} is approximately
\beq
t_\text{orb.} \simeq \pi \, G \, M_\odot (- m_\x / 2 E)^{3/2}
~.
\eeq
Also in this limit, the total orbital energy of a bound particle at $r$ is negligible compared to the gravitational potential energy at production; the magnitude of $E$ is bounded by $\abs{E}<m_\x \abs{\Phi(r)}\ll m_\x \abs{\Phi(r^\p)}$. With these approximations, the integral over $E$ and $v_\rho$ in \Eq{eq:nexact1} are analytically tractable, such that
\beq
\label{eq:napprox1}
n \simeq 2^{\ind-\frac{5}{2}} \, \frac{3 \, G  \, M_\odot \, t}{m_\x \, r^4} \int d^3 \xv^\p ~ \Qv (\xv^\p)  \, |\Phi (\xv^\p)|^{\ind-\frac{1}{2}}
~.
\eeq
We note that \Eq{eq:napprox1} agrees with the limiting form given in Refs.~\cite{VanTilburg:2020jvl,Lasenby:2020goo}.\footnote{Note that these references use a slightly different convention for the production rate, $\widetilde{Q} = 2^{3/2} \pi \, \Qv$.} We have numerically checked that the approximate expression in \Eq{eq:napprox1} underpredicts the local density at Earth as determined from \Eq{eq:nexact1} by at most a couple percent, making it both accurate and conservative. We emphasize that the $1/r^4$ scaling of \Eq{eq:napprox1} is only valid outside the Sun, and that the profile is flattened inside the Sun; this means that the integrated number of MCPs produced is finite and does not diverge as $r \to 0$. Note that this flattening of the density profile is purely a consequence of classical orbital dynamics.

From the derivation outlined above, we can glean several insights. First, the dependence on the velocity power law index $\ind$ (introduced in \Eq{eq:Qvdef}) primarily enters in the spatial integral of \Eq{eq:napprox1}. Since the solar potential is $\mathcal{O}(10^{-5})$ inside the Sun (see the right panel of \Fig{fig:Sunstuff}), production rates for processes with larger values of $\ind$ are suppressed. This can be interpreted as being due to the fact that processes whose rates have steeper velocity scalings are more penalized by the requirement that the velocity of the particle not exceed the solar escape velocity. We also see that compared to an unbound flux, which has a geometrical $1/r^2$ density profile, the bound population is more centrally concentrated with a $1/r^4$ profile due to the steepness of the gravitational potential closer to the Sun. Finally, from \Eq{eq:nexact1}, we see that the integral over energy $E$ has the most support when $E \sim m_\x \, \Phi(r)$, indicating that at a given radius $r$ the density of bound particles is dominated by those near their aphelia with velocities very close to zero. This is a direct consequence of the fact that bodies orbiting the Sun spend most of their time near their aphelia.

\subsection{Phase Space Saturation from Absorption}
\label{sec:saturation}

In the previous subsection, we assumed that the particles emitted in the solar interior are not reabsorbed, such that the phase space density of \Eq{eq:f1} continues to grow linearly with time $t$. This is valid provided that $f_p \ll 1$, where $f_p$ is the \emph{momentum} phase space density, which in the non-relativistic limit is related to the velocity phase space density $f$ of \Eq{eq:f1} by
\beq
\label{eq:fp}
f_p (r) \simeq (1/g_\text{spin}) \, (2 \pi / m_\x)^3 \, f(r, v_r, v_\theta)
~,
\eeq
where $g_\text{spin}$ is the number of internal spin degrees of freedom. 

However, at sufficiently large densities, the absorption of MCPs in the Sun can balance the rate of production, preventing further density growth. For instance, as discussed in Refs.~\cite{VanTilburg:2020jvl,Lasenby:2020goo}, for a solar basin consisting of axions or dark photons, detailed balance implies that this saturation point occurs once $f_p \sim f_\text{eq.}$, where $f_\text{eq.}$ is the Bose-Einstein equilibrium distribution at solar temperature $T_\odot \sim 1 \ \keV$. However, as we now discuss, this argument only applies to processes in which a single basin particle is emitted and absorbed by interacting with the stellar environment and hence cannot be applied straightforwardly to models where the emission of a single dark sector particle is forbidden, as is the case for MCPs.

The dominant production process arises from the decay of a solar plasmon into a pair of MCPs, $\g^* \to \text{MCP} ~ \text{MCP}$. Depending on the MCP mass, one or both of the outgoing states can be emitted non-relativistically with a velocity below the solar escape velocity. Energy-momentum conservation implies that the decay of a plasmon into two non-relativistic MCPs only occurs when the energy of the plasmon is nearly equal to twice the MCP mass (i.e., just above threshold). Plasmons only have this energy for a narrow range of electron densities, corresponding to a thin spherical shell within the solar volume. As a result, the rate for pair-producing MCPs that both contribute to the solar basin density is parametrically suppressed (by the volume ratio of the thin shell and the total solar volume) compared to decays that produce one non-relativistic and one relativistic particle in the final state. As we now show, this implies that $f_p$ saturates very close to unity for fermionic MCPs and well above unity for bosonic MCPs. 

To see this explicitly, consider the evolution of the basin phase space density $f_p$ as described by the Boltzmann equation, which schematically is of the form 
\begin{align}
\label{eq:satBoltz}
\dot{f}_p &\sim \, \mathcal{C}_\text{prod.}^{(1)} \, (1 \pm f_p) -  \mathcal{C}_\text{abs.}^{(1)} \, f_p 
\nl
& \, \, +  \mathcal{C}_\text{prod.}^{(2)} \, (1 \pm f_p)^2 -  \mathcal{C}_\text{abs.}^{(2)} \, f_p^2
~,
\end{align}
where the dot denotes a derivative with respect to time and $\pm$ correspond to bosonic/fermionic MCPs. Above, $\mathcal{C}^{(1,2)}_\text{prod.}$ and $\mathcal{C}^{(1,2)}_\text{abs.}$ encapsulate all the factors dictating production or absorption that do not directly depend on $f_p$, where the superscripts $(1,2)$ denote whether the velocities of one or both of the MCPs are below the solar escape velocity. In the case that only a single MCP is non-relativistic, the other one is relativistic, which we denote with a phase space density $f_p^\p$. 

\Eq{eq:satBoltz} can be simplified by noting that production requires an initial state thermal plasmon, $\mathcal{C}_\text{prod.}^{(1)} \propto f_{\g^*}$, where the plasmon density $f_{\g^*}$ is described by a Bose-Einstein equilibrium distribution at temperature $T_\odot$. On the other hand, the absorption coefficient $\mathcal{C}_\text{abs.}^{(1)}$ is suppressed by the small phase space of the relativistic solar flux of MCPs,   $\mathcal{C}_\text{abs.}^{(1)} \propto f_p^\p \ll f_{\g^*}$. As a result, the absorption of one non-relativistic MCP along with one relativistic MCP is subdominant to the time-reversed process, $\mathcal{C}_\text{abs.}^{(1)} \ll \mathcal{C}_\text{prod.}^{(1)}$. For the remaining terms in \Eq{eq:satBoltz} corresponding to two non-relativistic MCPs, we can factor out the plasmon phase space density, in which case detailed balance yields
\beq
\frac{\mathcal{C}_\text{prod.}^{(2)}}{ \mathcal{C}_\text{abs.}^{(2)}} = \frac{f_{\g^*} (2 m_\x)}{1 + f_{\g^*} (2m_\x)}
~,
\eeq
where $f_{\g^*}$ is evaluated at an energy of twice the MCP mass. With these simplifications, the saturation density of the MCP solar basin is determined by setting $\dot{f}_p = 0$ in \Eq{eq:satBoltz} and solving for $f_p$. We find that $f_p$ saturates near 
\beq
f_p (\text{boson}) \simeq \frac{\mathcal{C}_\text{prod.}^{(1)}}{\mathcal{C}_\text{prod.}^{(2)}} ~  f_{\g^*} (2 m_\x)
\gg 1
\eeq
for bosonic MCPs and 
\beq
f_p  (\text{fermion}) \simeq 1- \frac{\mathcal{C}_\text{prod.}^{(2)}}{\mathcal{C}_\text{prod.}^{(1)}} ~ \frac{1+ f_{\g^*} (2 m_\x)}{f_{\g^*} (2 m_\x)} \simeq 1
\eeq
for fermionic MCPs, where we used that production of a single non-relativistic MCP is much more likely than the emission of two non-relativistic MCPs ($\mathcal{C}_\text{prod.}^{(1)} \gg \mathcal{C}_\text{prod.}^{(2)}$).

The above argument demonstrates that unlike models of singly-produced basin particles, as considered in Refs.~\cite{VanTilburg:2020jvl,Lasenby:2020goo}, the stellar basin density of \emph{pair-produced} particles saturates at much larger values, i.e., once the phase space occupancy $f_p$ is nearly or highly degenerate, for fermions or bosons, respectively. We note that for bosonic MCPs, the timescale to obtain $f_p \gtrsim 1$ can be much shorter than the age of the solar system, at which point the density begins exponentially growing to its final saturated value. A precise calculation of the rate of exponential growth and the saturation density for bosonic MCPs is beyond the scope of this paper. We leave a detailed exploration to future work~\cite{KVT}. For the remainder of this study, we therefore mostly focus on the case of fermionic MCPs and self-consistently incorporate the effect of solar absorption by imposing the upper bound $f_p \lesssim 1$. The corresponding upper bound on the number density $n$ can be evaluated by integrating over the basin phase space. As noted in the previous subsection, the occupied phase space is limited to lie within the interval $|v_{r, \rho}| \lesssim v_{r, \rho}^\text{max.}$ where $v_r^\text{max.} \simeq v_\text{esc.} (r)$ and $v_\rho^\text{max.} \sim (r^\p / r) \, v_\text{esc.} (r^\p)$ (see \Eq{eq:vrhomax}). Taking $f_p \simeq 1$ and four spin degrees of freedom, the integrated number density for fermionic MCPs is thus bounded by $n \lesssim n_\text{sat.}$, where the saturation density is
\begin{align}
\label{eq:satunpert}
n_\text{sat.} (r) &\simeq \frac{m_\x^3}{\pi^2} \, v_\text{esc.} (r) \, \left\langle (v_\rho^{\text{max.}})^2 \right\rangle_{V_\odot} \propto \frac{1}{r^{5/2}}
~,
\end{align}
and the angle brackets denote an average over the solar volume, weighted by the differential production rate $dn / d^3 \xv^\p$.

We conclude this section by briefly commenting on a possible modification to the arguments detailed above. In particular, our analysis has assumed that the basin phase space from solar production remains unperturbed over the solar lifetime. However, as discussed in Refs.~\cite{VanTilburg:2020jvl,Lasenby:2020goo}, it is possible that gravitational interactions with other planets, such as Earth and Jupiter, may perturb the orbits of MCPs on timescales shorter than the age of the solar system. Such interactions can significantly modify the phase space density of  \Sec{sec:phasespaceproduction} for aphelia larger than $r_\text{max.} \sim 1 \ \text{AU}$. We do not incorporate a careful analysis of such effects since this requires a dedicated numerical simulation, but instead follow the approach of Refs.~\cite{VanTilburg:2020jvl,Lasenby:2020goo} by bracketing the range of possible effects from planetary encounters. For instance, if the initial radial orbits are completely isotropized due to gravitational interactions, then we can determine the maximum density by integrating $f_p \simeq 1$ but now over $0 \lesssim v \lesssim v_\text{esc.} (r)$, such that
\beq
\label{eq:satpert}
n_\text{sat.} (r) \simeq \frac{2 m_\x^3}{3 \pi^2} \, v_\text{esc.}(r)^3 \propto \frac{1}{r^{3/2}}
~.
\eeq
Note that the radial profile of the saturation density $n_\text{sat.}$ in either \Eq{eq:satunpert} or \Eq{eq:satpert} differs from the profile of the unsaturated density $n \propto r^{-4}$ in \Eq{eq:napprox1}.

\section{Production Inside the Sun}
\label{sec:solarproduction}

In this section, we present the results for the solar production rate $\Qv$ of non-relativistic MCPs with power-law index $\ind = 1$ (see \Eq{eq:Qvdef}), which is needed in Eqs.~(\ref{eq:f1}) and (\ref{eq:nexact1}) to calculate the phase space and number density of the solar basin. MCPs lighter than the typical plasma frequency in the solar interior, $\w_p \sim 100 \ \eV$, can be produced efficiently. Such MCPs are dominantly produced from the decay of plasmons, which are electromagnetic excitations in the plasma with altered in-medium dispersion relations and polarization vectors compared to the photon in vacuum~\cite{Davidson:2000hf,Vinyoles:2015khy}. Plasmon decay into particle-antiparticle pairs is the dominant production channel since this process is lower order in $\alpha_\text{em}$ compared to, e.g., Compton or bremsstrahlung-like reactions, is lower order in $q_\x$ than photon fusion, and is not suppressed by a low initial state number density as would be the case for, e.g., $e^+ e^-$ annihilation~\cite{Vinyoles:2015khy}.

\begin{figure}[t]
\includegraphics[width=\columnwidth]{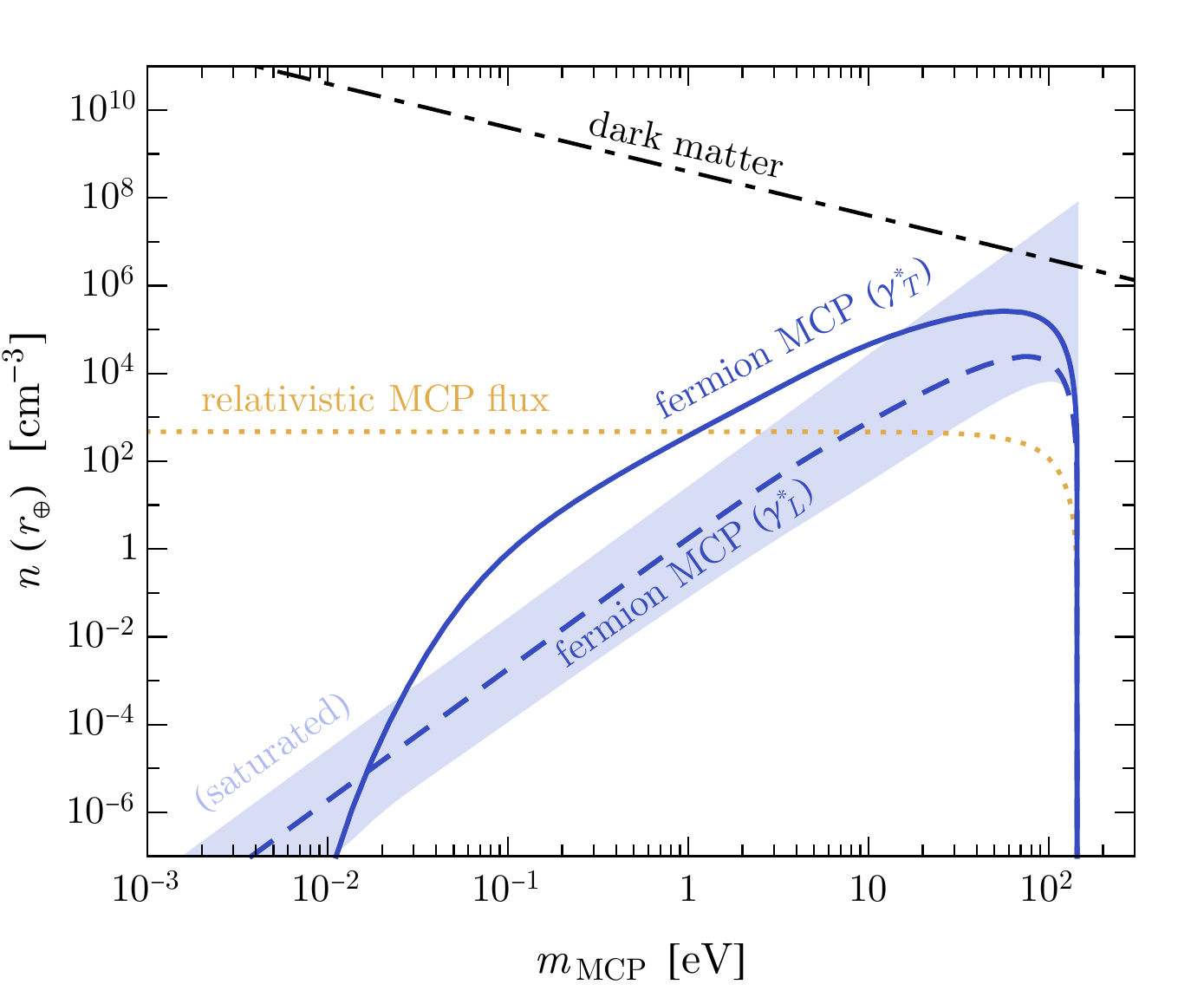}
\caption{The local number density $n (r_\oplus)$ of fermionic MCPs bound to the solar system as a function of the MCP mass $m_\x$, assuming the basin lifetime is comparable to the solar age of 4.5~Gyr. The MCP coupling is fixed to be $q_\x \simeq 2 \times 10^{-14}$, the largest value allowed by existing constraints for such masses~\cite{Davidson:2000hf,Vinyoles:2015khy}. The solid and dashed dark blue lines show the number density contributed by transverse and longitudinal plasmon decay according to \Eq{eq:napprox1}, which ignores Fermi statistics. The light blue band indicates the point at which the density saturates due to Fermi statistics; hence, the MCP density never actually exceeds this saturation density. The top and bottom of this band corresponds to a basin phase space that is maximally or minimally perturbed by gravitational encounters, respectively (see \Sec{sec:saturation}). For comparison, we also show the relativistic flux of solar emitted MCPs (dotted orange) and the local dark matter density assuming that it consists of particles of mass $m_\x$ (dot-dashed black).
}
\label{fig:production}
\end{figure}

 The rate for plasmon decay depends on the spin of the MCP (spin-0 or spin-1/2) and the polarization of the plasmon (transverse or longitudinal). Although our primary focus in this work is on fermionic MCPs, for completeness we also show the production rates for scalar MCPs below. In \App{app:MCPproduction}, we present the detailed derivations for each of these cases. The non-relativistic production rate for transverse plasmon decay to either fermionic or scalar MCPs is approximately
\begin{align}
\label{eq:QvT}
\Qv (\g^*_T) &\simeq \frac{\alpha_\text{em} \, q_\x^2}{4 \pi^3} \, m_\x \, \w_p^4 \, ~ f_{\g^*} (\w_p^2 / 2 m_\x)
\nl
&\times \begin{cases}
\big( 1 - \frac{4 m_\x^2}{\w_p^2} \big)^{1/2}  & (\text{fermion})
\\
\order (v^{\p \, 2}) & (\text{scalar})
~,
\end{cases}
\end{align}
where $f_{\g^*} (\w_{\g^*}) = 1 / (e^{\w_{\g^*} / T_\odot} - 1)$ is the plasmon phase space distribution. Note that transverse plasmon decay to scalar MCPs is parametrically suppressed by the small MCP velocity, analogous to the well-known $p$-wave suppression in the non-relativistic limit of spin-1 mediated annihilations of scalar particles~\cite{Kumar:2013iva}. Alternatively, the decay rate for longitudinal plasmons is given by
\begin{align}
\label{eq:QvL}
\Qv (\g^*_L) &\simeq \frac{\alpha_\text{em} \, q_\x^2}{4 \pi^3} \, m_\x^3 \, \w_p \, f_{\g^*} (\w_p)
\nl
&\times \begin{cases}
2 \, m_\x \, \big( 1 - \frac{2 m_\x}{\w_p} \big)^{1/2} &(\text{fermion})
\\
\frac{1}{2} \, \w_p \, \big( 1 - \frac{2 m_\x}{\w_p} \big)^{3/2}   &(\text{scalar})
~.
\end{cases}
\end{align}

From Eqs.~(\ref{eq:QvT}) and (\ref{eq:QvL}) we take note of a few important insights. First, for scalar MCPs, longitudinal plasmon decay always dominates over transvere plasmon decay, due to the different $v^\p$ dependence. Second, for fermionic MCPs, the hierarchy between the different plasmon polarizations depends on the particular value of the mass; for fermionic MCP masses near the kinematic threshold, $m_\x \sim \w_p / 2$, the decay of transverse plasmons is slightly enhanced compared to longitudinal plasmon decays due to the larger power of $\w_p$ in the former case. However, due to the argument of the plasmon phase space distribution $f_{\g^*}$ in \Eq{eq:QvT}, transverse plasmon decay to fermionic MCPs is exponentially suppressed compared to the decay of longitudinal plasmons for masses well below the kinematic threshold, $m_\x \ll \w_p^2/T_\odot$.

 The results for the terrestrial number density of fermionic MCPs are shown in \Fig{fig:production} as a function of the MCP mass and the coupling fixed to $q_\x \simeq 2 \times 10^{-14}$, the largest allowed by existing constraints in this mass range~\cite{Davidson:2000hf,Vinyoles:2015khy}. The bound density can exceed the relativistic flux of emitted MCPs by orders of magnitude for $m_\x \sim 100 \ \eV$.

\section{Direct Deflection}
\label{sec:deflection}

So far, we have discussed the formation of a gravitationally bound population of MCPs. In this section, we focus on the prospects for detecting this population. Direct deflection was recently proposed in Ref.~\cite{Berlin:2019uco} as a new technique to detect feebly interacting ambient particles that couple to long-ranged forces, such as electromagnetism. Although this setup was originally introduced to detect sub-GeV dark matter, it also straightforwardly applies  to the gravitationally bound population of MCPs that is discussed in this work. 

A basic schematic of the setup is shown in Fig.~\ref{fig:cartoon}. 
An approximately spatially uniform charge-symmetric MCP population passes into a shielded region, with an electric field $E_\text{def.}$ oscillating at angular frequency $\w$. This region is referred to as the “deflector.” Inside the deflector, the MCPs are subject to an electric force that slightly separates positively and negatively charged particles, resulting in oscillating charge densities $\rho_\pm$ that propagate out of the deflector and into a spatially distinct shielded detection region. In the detector, these MCP charge densities induce a small oscillating electromagnetic field $E_\text{sig.}$ at the same frequency $\w$, which can be measured using an electric field pickup antenna coupled to a resonant LC circuit tuned to the same frequency. 

The detailed spatial dependence of $\rho_\pm$ is directly sensitive to the MCP velocity distribution. However, the typical magnitude of the electric field signal is largely independent of these considerations, provided that the deflector and detector regions are not too far spatially separated. In particular, for an optimally configured experimental setup, these millicharge densities source an electric field that scales according to Eqs.~(\ref{eq:approxcharge}) and (\ref{eq:approxEsig}) as 
\beq
\label{eq:EsigGeneric}
E_\text{sig.} \sim m_{D, \x}^2 \, \varphi_\text{def.} \, R_\text{def.} \, e^{i \w t}
~,
\eeq
where $\varphi_\text{def.}$ and $R_\text{def.}$ are the electric potential and spatial size of the deflector, respectively, $m_{D, \x} \simeq e q_\x \sqrt{n (r_\oplus) / T_\x}$ is the MCP contribution to the photon Debye mass, and $T_\x \simeq m_\x \, \langle v^2 \rangle / 3$ is the effective MCP temperature.

In \Sec{sec:chargedensity}, we give an overview of the basic formalism needed to precisely calculate the deflector-induced MCP charge density $\rho_\pm$. Additional details are provided in \App{app:chargedensity}. As shown below in \Eq{eq:deflectiongeneral}, $\rho_\pm$ is directly related to the MCP velocity distribution in the laboratory frame. We evaluate the charge density for a basin phase space that is either minimally or maximally perturbed by gravitational interactions in Secs.~\ref{sec:unperturbcharge} and \ref{sec:perturbcharge}, respectively. We find that the signal electric field $E_\text{sig.}$ roughly matches the parametric form of \Eq{eq:EsigGeneric} in either case. 

\subsection{Millicharge Overdensities}
\label{sec:chargedensity}

In the following subsections, we adopt the notation where vectors with tildes, such as $\tilde{\xv}$, denote positions with respect to the center of the deflector apparatus, in order to differentiate from the heliocentric notation of the previous sections. As derived in Ref.~\cite{Berlin:2019uco}, the induced MCP charge overdensity at position $\tilde{\xv}$ with respect to the center of the deflector is approximately
\begin{align}
\label{eq:deflectiongeneral}
&\rho_\pm (\tilde{\xv}, t) \simeq - \, \frac{(e q_\x)^2}{m_\x} ~ e^{i \w t} 
\nl
&\times \int d v ~ \int_{V_\text{def.}} \hspace{-0.2cm} d^3 \tilde{\xv}^\p ~  f(r_\oplus, v\, \hat{\vv} + \vv_\oplus) ~ \frac{\rho_\text{def.} (\tilde{\xv}^\p)}{|\tilde{\xv} - \tilde{\xv}^\p|}
~,
\end{align}
where $\rho_\text{def.}$ is the amplitude of the oscillating deflector charge density (which drives the electric field $E_\text{def.}$), the integral over $\tilde{\xv}^\p$ is over the volume of the deflector $V_\text{def.}$, $\vv_\oplus \simeq - 10^{-4} \, \hat{\tilde{{\xv}}}$ is Earth's orbital velocity, and we have defined the unit vector $\hat{\vv} \equiv (\tilde{\xv} - \tilde{\xv}^\p) / |\tilde{\xv} - \tilde{\xv}^\p|$. The integral over the basin velocity $v$ includes the velocity distribution $f(r, \vv)$ of \Eq{eq:f1} boosted to Earth's frame. For concreteness, we will consider a deflector as depicted in \Fig{fig:cartoon}, consisting of an oscillating  point charge $Q_\text{def.} \, e^{i \w t}$ located at the origin $|\tilde{\xv}| = 0$  surrounded by a grounded spherical shield of radius $R_\text{def.}$, which possesses a corresponding surface charge. This deflector charge density configuration is thus expressed as the sum of these two components, i.e.,
\beq
\label{eq:rhodefpoint}
\rho_\text{def.} (\tilde{\xv}) \simeq Q_\text{def.} \, \Big( \delta^3(\tilde{\xv}) - \frac{1}{4 \pi R_\text{def.}^2} \, \delta (|\tilde{\xv}| - R_\text{def.})   \Big)
~,
\eeq
such that the total integrated charge of the deflector and shield is zero.

As discussed in Ref.~\cite{Berlin:2019uco}, the derivation of \Eq{eq:deflectiongeneral} assumes that the period of a deflector oscillation ($\sim 1 / \w$) is long compared to the time it takes for a typical MCP to traverse the deflector, i.e., $\w \ll v_\oplus / R_\text{def.}$. Throughout, we will work in this limit because for $\w \gtrsim v_\oplus / R_\text{def.}$ the induced millicharge overdensity $\rho_\pm$ is parametrically suppressed, as it averages out to zero when the oscillations are too rapid. \Eq{eq:deflectiongeneral} also relies on various other assumptions; in addition to the previous criterion, the above expression holds provided that the MCP population can be treated as a continuum (i.e., many particles per experimental volume so that Poisson fluctuations can be ignored) that is weakly perturbed by the deflector\footnote{We note that the signal persists for couplings larger than this perturbative limit, although the calculation is less tractable since it cannot be estimated to leading order in perturbation theory, as was done in Ref.~\cite{Berlin:2019uco}. For the experimental parameters adopted in this work, the perturbative criteria holds for most of the parameter space shown in \Fig{fig:reach}, except for a small region  corresponding to $q_\x \gtrsim 10^{-14} \times (m_\x / \text{eV})$. We note, however, that since the reach only mildly depends on the strength of the deflector electric field as $q_\x \propto E_\text{def.}^{-1/4}$ (see \Sec{sec:experiment}), the region over which this weak-coupling approximation holds could be enlarged without significantly effecting the projected sensitivity.} ($e q_\x \varphi_\text{def.} \ll T_\x$) and that backreactions effects from self-interactions are negligible~\cite{Berlin:2019uco}. The latter criterion requires that the MCPs do not screen the deflector charge density $\rho_\text{def.}$ on length-scales smaller than the deflector itself, i.e., $(m_D^\p)^{-1} \gg R_\text{def.}$, where $m_D^\p \simeq \sqrt{4 \pi \alpha^\p \, n (r_\oplus) / T_\x}$ is the dark photon Debye mass. For the range of $\alpha^\p$ considered in this work, this is satisfied (see \Sec{sec:interactions}). 

We mentioned above that this formalism relies on approximating the MCP population as a continuum whose properties depend on orbital dynamics and on the deflection from the experimental apparatus. However, for sufficiently light MCPs, quantum mechanical effects are important once the de Broglie wavelength of the MCP particle becomes macroscopic. In order to investigate when such a transition between classical and quantum behavior occurs, we note that the Euler equation of classical fluid mechanics can be applied to the quantum regime if an additional pressure term is included, in the form of the Madelung equation (see, e.g., Ref.~\cite{Spiegel:1980ykb}). The analysis above is unmodified provided that the force from this quantum pressure term is negligible compared to the electromagnetic force of the deflector, i.e., $e q_\x \, m_\x \, E_\text{def.} \gg 1/R_\text{def.}^3 \, $. For a driven deflector of field strength $E_\text{def.} \sim 10 \ \text{kV} / \cm$ over a region $R_\text{def.} \sim 1 \ \text{m}$, the classical limit corresponds to $m_\x \gg \text{meV} \times ( 10^{-17} / q_\x)$. As we will show in \Sec{sec:experiment}, the proposed deflection setup is sensitive to MCP masses and couplings that do indeed reside in this classical regime. 

Furthermore, the effect of Pauli-blocking does not enter into the formalism above, even though for sufficiently large couplings the phase space occupancy of fermionic MCPs is expected to be nearly degenerate (as discussed in  \Sec{sec:saturation}). This is a consequence of time-reversal symmetry, which implies that two MCPS from initially distinct regions of phase space cannot propagate along trajectories such that they end up at the same point in phase space after passing through the electric field of the deflector; if this was possible, then the time-reversed classical trajectory would not be deterministic.

As an aside, note that for a MCP velocity distribution that is isotropic in Earth's frame (such as a lab-frame Maxwellian distribution), then \Eq{eq:deflectiongeneral} simplifies to
\beq
\label{eq:deflectionhomogenous}
\rho_\pm (\tilde{\xv}, t) \simeq - m_{D, \x}^2 ~ \varphi_\text{def.} (\tilde{\xv}) ~ e^{i \w t}
~~(\text{isotropic}),
\eeq
where $\varphi_\text{def.}$ is the electric potential of the deflector.\footnote{Note that \Eq{eq:deflectionhomogenous} is analogous to how a SM photon mass $m_\g$ modifies Gauss's law, i.e., $\grad \cdot \Ev = \rho - m_\g^2 \, \varphi$. In this sense, $m_{D,\x}$ is playing the role of the photon mass in the limit that the MCP velocity distribution is isotropic in the lab frame.} Note that this is \emph{not} the case here;  the MCP solar basin does not corotate with Earth's orbit, and so the velocity distribution is anisotropic in the lab frame. Regardless of this difference, motivated by the form of \Eq{eq:deflectionhomogenous}, we define the constant
\beq
\label{eq:rhoDebye}
\rho_\pm^{(\text{Debye})} \equiv - \frac{(e q_\x)^2 \, n (r_\oplus)}{m_\x v_\oplus^2} ~ \frac{Q_\text{def.}}{4 \pi R_\text{def.}}
~,
\eeq
which serves as a useful comparison to our numerical results in Secs.~\ref{sec:unperturbcharge} and \ref{sec:perturbcharge} below. 

\subsection{Unperturbed Velocity Distribution}
\label{sec:unperturbcharge}

In this and the following subsections, we explicitly evaluate the induced MCP charge density $\rho_\pm$ for two different basin phase space distributions. We choose a coordinate orientation such that the axis running from the Sun to the Earth is along the $+\tilde{z}$ direction and the orbital motion of the Earth is chosen to lie along the $- \tilde{x}$ direction.

\begin{figure}[t]
\includegraphics[width=\columnwidth]{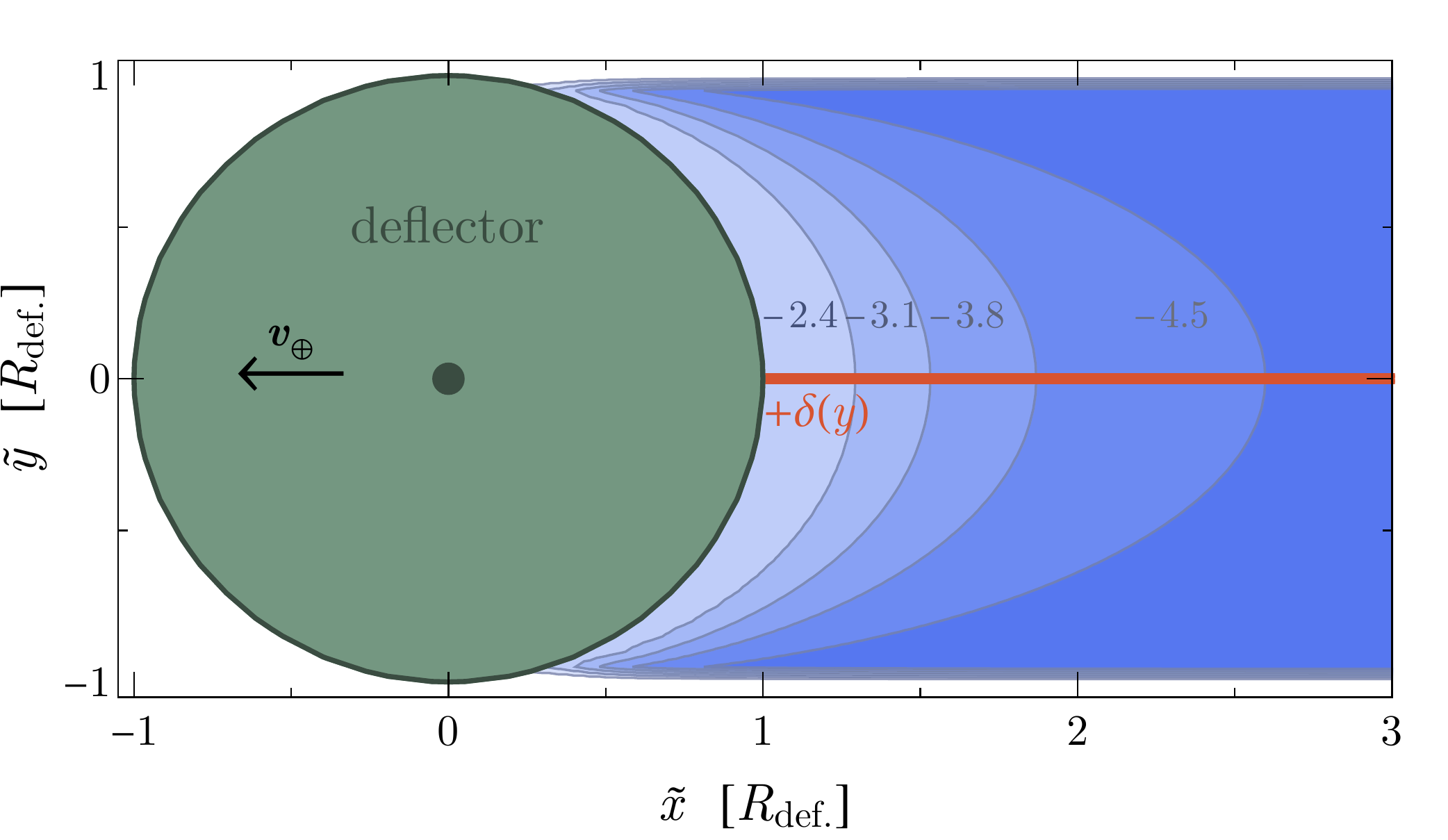}
\caption{The amplitude of the MCP charge density $\rho_\pm (\tilde{\xv})$ (red and blue regions) in the $\tilde{x}-\tilde{y}$ plane at $\tilde{z} = 0$, in units of $\rho_\pm^{(\text{Debye})}$ (see \Eq{eq:rhoDebye}), for a MCP velocity distribution that is unperturbed by gravitational interactions (see \Eq{eq:unpertf}). The green region is the shielded deflector, with a driven charge configuration consisting of a point charge and grounded spherical shell. The heliocentric radial direction is along the $\tilde{z}$-axis and the orbital motion of the Earth is along the negative $\tilde{x}$-axis. The point charge of the deflector induces the red region, which consists of a MCP surface charge density at $\tilde{y} = 0$, as in \Eq{eq:surfacecharge}. The spherical shell of the deflector induces the oppositely charged blue regions, as in \Eq{eq:shellcharge}.
}
\label{fig:xydeflect0}
\end{figure}

To begin, let us assume that the basin phase space is unperturbed by gravitational or hidden sector interactions, such that \Eq{eq:f1} accurately describes the local MCP velocity distribution. We can gain some analytic insight by first roughly approximating \Eq{eq:f1} as a purely radial distribution (with no velocity support in the transverse directions) that is uniformly flat below the local escape velocity, so that the velocity distribution in the solar frame is 
\beq
\label{eq:unpertf}
f(r_\oplus, \vv) \simeq n (r_\oplus) ~ \frac{\delta(v_x) \, \delta(v_y)}{2 v_\text{esc.}(r_\oplus)} ~ \Theta(v_\text{esc.}(r_\oplus) - |v_z|)
~.
\eeq
Using \Eq{eq:unpertf} in \Eq{eq:deflectiongeneral}, the millicharge overdensity $\rho_\pm$ can be evaluated analytically (see \App{app:chargedensity} for more details). \Fig{fig:xydeflect0} shows the total contribution to the amplitude of the MCP charge density $\rho_\pm (\tilde{\xv})$ in the $\tilde{x}-\tilde{y}$ plane at $\tilde{z} = 0$, normalized by $\rho_\pm^{(\text{Debye})}$ as defined in \Eq{eq:rhoDebye}.

\begin{figure*}[t]
\includegraphics[width= 0.95 \columnwidth]{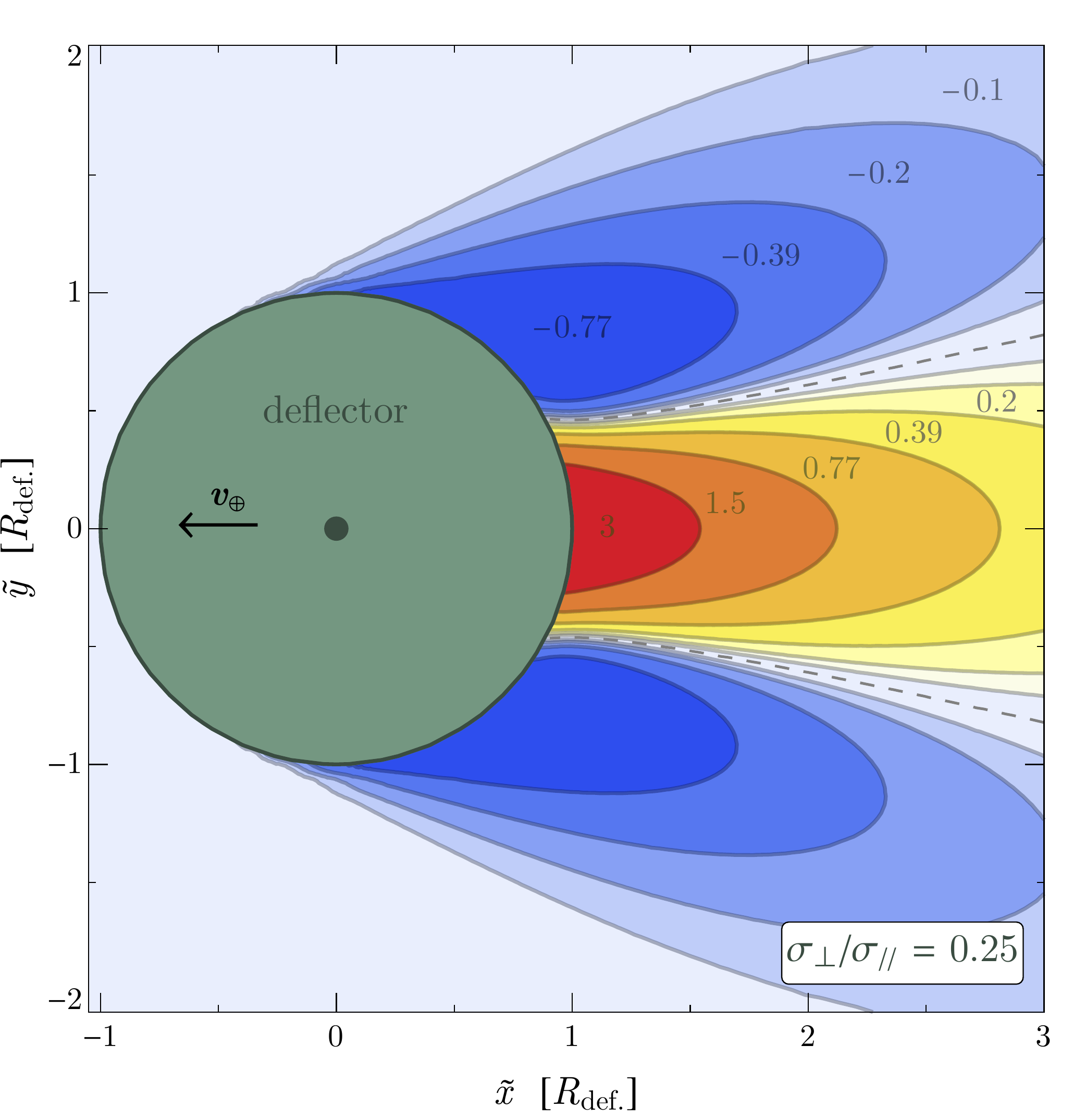}
\hspace{1cm}
\includegraphics[width= 0.95 \columnwidth]{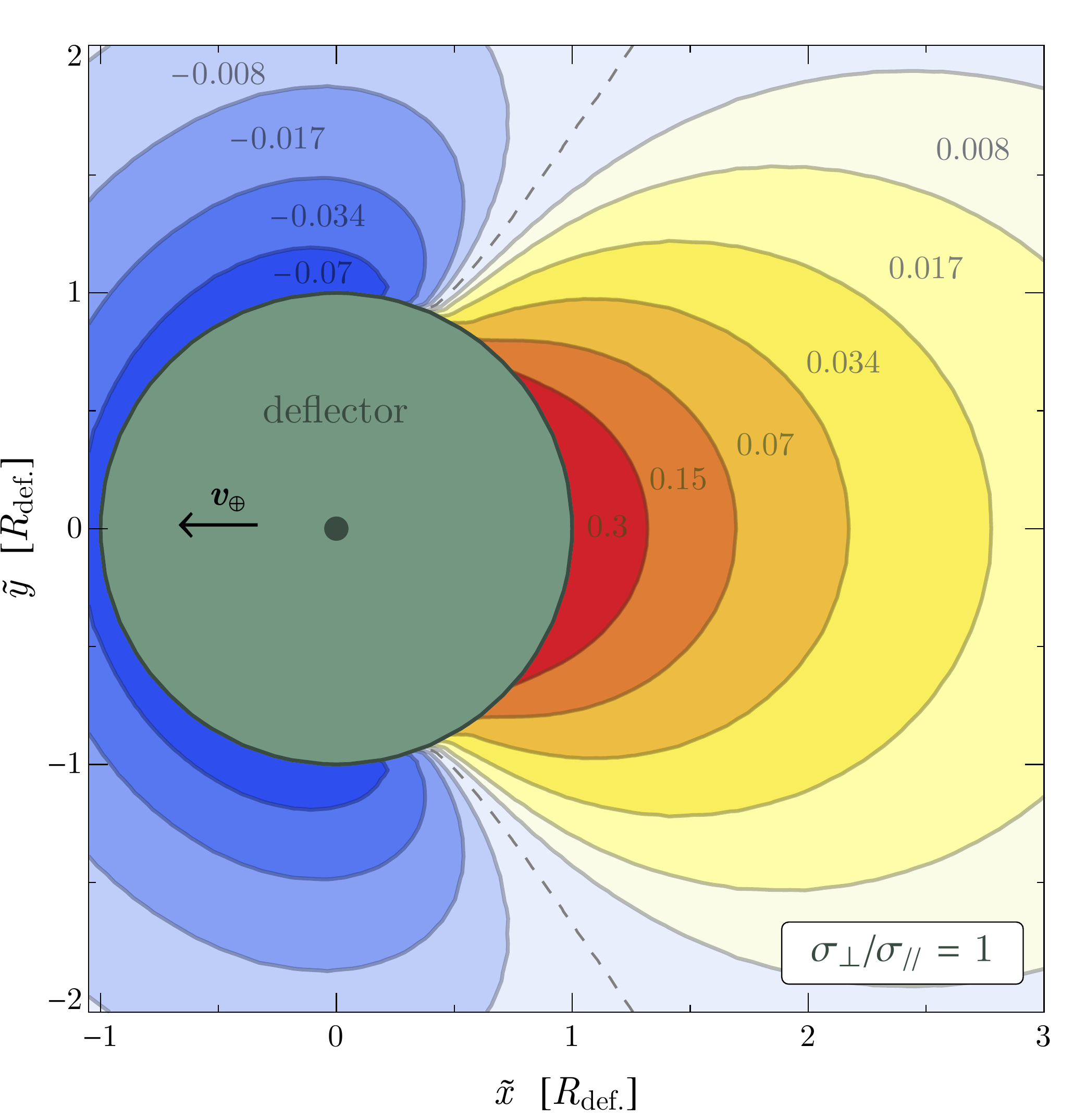}
\caption{The amplitude of the MCP charge density $\rho_\pm (\tilde{\xv})$ (red/orange/yellow and blue regions) in the $\tilde{x}-\tilde{y}$ plane at $\tilde{z} = 0$, in units of $\rho_\pm^{(\text{Debye})}$ (see \Eq{eq:rhoDebye}), for a MCP velocity distribution that is slightly (left panel) or maximally (right panel) perturbed by gravitational encounters, corresponding to relative velocity dispersions of $\sigma_\perp / \sigma_{\para}  = 0.25$ and $\sigma_\perp / \sigma_{\para}  = 1$, respectively, fixing $\sigma_{\para} = v_\text{esc.} (r_\oplus)$ (see \Eq{eq:gvperturb}). The green region is the shielded deflector, with a driven charge configuration consisting of a point charge and grounded spherical shell. The heliocentric radial direction is along the $\tilde{z}$-axis and the orbital motion of the Earth is along the negative $\tilde{x}$-axis. The red/orange/yellow and blue regions correspond to positive and negative values of $\rho_\pm (\tilde{\xv}) / \rho_\pm^{(\text{Debye})}$, respectively. Along the dashed gray lines, $\rho_\pm (\tilde{\xv})$ crosses through zero.
}
\label{fig:xydeflect}
\end{figure*}

We find that the driven point charge $Q_\text{def.}$ of the deflector (corresponding to the first term of \Eq{eq:rhodefpoint}) induces a MCP surface charge density in the region $\tilde{y} = 0$ and $\tilde{x} \gtrsim |\tilde{z}|$ that is parametrically of size
\beq
\label{eq:surfacecharge}
\sigma_\pm^{(\text{point})} \sim - \frac{(e q_\x)^2 \, n (r_\oplus)}{2 m_\x \, v_\oplus^2} ~ Q_\text{def.}
~.
\eeq
This MCP surface charge density is shown as the red region of \Fig{fig:xydeflect0}. This component of the induced MCP charge density sources an electric field $E \sim \sigma_\pm^{(\text{point})}$, in agreement with \Eq{eq:EsigGeneric}. 

The driven spherical shell of the deflector (corresponding to the second term of \Eq{eq:rhodefpoint}) induces a MCP charge density in the region $\tilde{x} >0$ and $|\tilde{y}| \leq R_\text{def.}$ that is parametrically of size
\beq
\label{eq:shellcharge}
\rho_\pm^{(\text{shell})} \sim \frac{(e q_\x)^2 \, n (r_\oplus)}{4 m_\x \, v_\oplus^2} ~ \frac{Q_\text{def.}}{R_\text{def.}}
~.
\eeq
This contribution to the MCP charge density is shown as the blue region of \Fig{fig:xydeflect0}. A uniform charge density of this magnitude over a length-scale comparable to the deflector size sources an electric field $E \sim \rho_\pm^{(\text{shell})} \, R_\text{def.}$. Although this contribution is of opposite sign compared to that of $\sigma_\pm^{(\text{point})}$, the distinct spatial dependence of $\sigma_\pm^{(\text{point})}$ and $\rho_\pm^{(\text{shell})}$ implies that a shielded detector could be placed in a particular location to measure either contribution individually. Furthermore, since this is only a mild cancellation, in most regions the total induced MCP charge density sources an electric field comparable in magnitude to $E_\text{sig.} \sim |\sigma_\pm^{(\text{point})}| \sim | \rho_\pm^{(\text{shell})} | \, R_\text{def.} \sim (e q_\x)^2 \, n (r_\oplus)  \, Q_\text{def.} / (m_\x \, v_\oplus^2)$, consistent with the generic expectation of \Eq{eq:EsigGeneric}. 

These results relied on approximating the basin velocity distribution using the simple delta function and top-hat parameterization of \Eq{eq:unpertf}. This vastly simplified the evaluation of the integrals needed to compute the MCP charge density in \Eq{eq:deflectiongeneral}. We have explicitly checked numerically that these approximations accurately describe to within $\mathcal{O}(10) \%$ the MCP charge densities using the exact form of the unperturbed velocity distribution in \Eq{eq:f1}. 

\subsection{Perturbed Velocity Distribution}
\label{sec:perturbcharge}

As mentioned previously, perturbations to the MCP phase space may arise via gravitational perturbations from planetary encounters. If the timescales associated with these effects are short compared to the age of the solar system, the velocity distribution at Earth is expected to differ qualitatively from \Eq{eq:f1}. For a strongly perturbed velocity distribution, we model it as a Gaussian in the solar frame, 
\beq
\label{eq:gvperturb}
f(r_\oplus, \vv) \simeq \frac{n (r_\oplus)}{\pi^{3/2} \, \sigma_\perp^2 \, \sigma_{\para}} ~ e^{-(v_x^2 + v_y^2) / \sigma_\perp^2} ~ e^{-v_z^2 / \sigma_{\para}^2}
~,
\eeq
where the velocity dispersions $\sigma_\perp$ and $\sigma_{\para}$ in the direction perpendicular ($\tilde{x}$ and $\tilde{y}$) and parallel ($\tilde{z}$) to the heliocentric radial direction, respectively, are taken to be independent. At the level of this Gaussian approximation, we have ignored the fact that \Eq{eq:gvperturb} unphysically populates phase space above the solar escape velocity $v_\text{esc.} (r_\oplus) \sim 10^{-4}$. However, because the deflection signal is enhanced for the more slowly moving MCPs in the distribution, including particles in the tails of the Gaussian amounts to introducing a small error when calculating $\rho_\pm$. We will enforce that the dispersion satisfies $\sigma_\perp, \sigma_{\para} \lesssim v_\text{esc.} (r_\oplus)$ to ensure that this error is negligible.

To model a maximally perturbed distribution, we take the isotropic limit of \Eq{eq:gvperturb}, i.e., $\sigma_\perp \simeq \sigma_{\para} \simeq v_\text{esc.} (r_\oplus)$, whereas a less perturbed distribution is modelled with a smaller perpendicular dispersion, $\sigma_\perp < \sigma_{\para} \simeq v_\text{esc.} (r_\oplus)$. We then numerically calculate the resulting MCP charge density using \Eq{eq:deflectiongeneral}. The results are shown for both a moderately perturbed and maximally perturbed velocity distribution in the left and right panel of  \Fig{fig:xydeflect}, respectively. As in \Fig{fig:xydeflect0}, $\rho_\pm (\tilde{\xv})$ is plotted in the $\tilde{x}-\tilde{y}$ plane and is normalized by $\rho_\pm^{(\text{Debye})}$. In the left and right panel of \Fig{fig:xydeflect}, we have set $\sigma_\perp / \sigma_{\para}$ to $0.25$ and $1$, respectively, fixing $\sigma_{\para} \simeq v_\text{esc.} (r_\oplus)$ in both cases. 

In the left panel of \Fig{fig:xydeflect}, the charge density distribution consists of a narrow positive region of $\rho_\pm (\tilde{\xv}) / \rho_\pm^{(\text{Debye})}$ (shown in shades of red, orange, and yellow) centered near the $\tilde{x}-\tilde{z}$ plane at $\tilde{y} = 0$, along with a region of negative $\rho_\pm (\tilde{\xv}) / \rho_\pm^{(\text{Debye})}$ (shown in shades of blue) away from $\tilde{y} = 0$ but still roughly within $-R_\text{def.} \lesssim \tilde{y} \lesssim R_\text{def.}$ for $0 < \tilde{x} \lesssim \text{few} \times R_\text{def.}$. This spatial dependence with respect to the deflector origin and the Earth velocity $\vv_\oplus$ is qualitatively similar to the maximally unperturbed case shown in \Fig{fig:xydeflect0}, which indicates that $\rho_\pm (\tilde{\xv})$ quickly approaches the unperturbed limit as $\sigma_\perp $ drops further below $\sigma_{\para}$. 

In the right panel of \Fig{fig:xydeflect}, the charge density distribution consists nearly entirely of a large positive region of $\rho_\pm (\tilde{\xv}) / \rho_\pm^{(\text{Debye})}$ for $\tilde{x} > 0$ surrounded by a region of negative $\rho_\pm (\tilde{\xv}) / \rho_\pm^{(\text{Debye})}$ that is much smaller in magnitude. Note that the magnitude of the charge density falls off much more rapidly as a function of $\tilde{r} \gg R_\text{def.}$ for $\sigma_\perp / \sigma_{\para}  = 1$ (right panel) than for  $\sigma_\perp / \sigma_{\para}  = 0.25$ (left panel). As discussed in Ref.~\cite{Berlin:2019uco}, this can be understood from the relative size of the velocity dispersion in either case, which can be thought of as ``diluting" the induced charge densities in the $\tilde{y} - \tilde{z}$ plane as they are ``dragged" downwind starting from the interior of the deflector shield to large positive $\tilde{x}$ ($\gg R_\text{def.}$). In both panels, the charge densities are of the expected size, i.e., $|\rho_\pm (\tilde{\xv}) / \rho_\pm^{(\text{Debye})}| \sim 10^{-1} - 10$, and hence source electric fields consistent with the generic expectation of \Eq{eq:EsigGeneric}. 

\subsection{Experimental Setup and Reach}
\label{sec:experiment}

The MCP charge densities discussed in the previous subsections oscillate at the frequency of the deflector $\w$ and propagate at a speed $v_\oplus$ in the $+ \tilde{x}$ direction~\cite{Berlin:2019uco}. The MCP constituents are extremely feebly coupled, with a mean free path many orders of magnitude larger than the size of the experimental setup. Thus, they can easily penetrate a shielded detector region placed ``downwind" of the deflector. For concreteness, we take the detector shield to be of the same size as the deflector but centered on the point $(\tilde{x}, \tilde{y} , \tilde{z}) = (2 R_\text{def.}, 0, 0)$ (referring to the coordinates of Figs.~\ref{fig:xydeflect0} and \ref{fig:xydeflect}). Upon penetrating the detector shield, the oscillating charge densities of MCPs source oscillating electromagnetic fields of the same frequency inside the quiet detection region. 

As mentioned above, this setup is assumed to operate in the quasi-static limit, $\w \lesssim v_\oplus / R_\text{def.} \simeq 30 \ \text{kHz} \times (1 \ \text{m}/ R_\text{def.})$, because this maximizes the signal strength. The optimal detector that can resonantly respond to such frequencies is an LC circuit because its resonant frequency is not directly dictated by its geometric size. The electric field sourced by the MCP charge density capacitively couples to the circuit, driving a small voltage at $\w$. If the resonant frequency of the circuit is tuned to the same frequency, then the small MCP-induced electric field rings up over many cycles inside the circuit, as quantified by the large quality factor, $Q_\text{LC} \gg 1$, of the detector. Similar technology has been implemented by the AURIGA experiment~\cite{Cerdonio:1997hz,Baggio:2005xp, Bonaldi:1998gcg, Bonaldi:1999mvu} to detect gravitational waves, which utilized a thermal-noise limited LC circuit operating at kHz frequencies with a quality factor of $Q_\text{LC} \sim 10^6$. This technology will be further developed by DM Radio~\cite{Silva-Feaver:2016qhh, Godfrey:2021tvs} to search for ultralight coherent bosonic dark matter; a future version of the existing prototype is expected to ultimately achieve an inductively-coupled thermal-noise limited setup with $Q_\text{LC} \sim 10^7$ and a detection volume of $V_\text{det.} \sim 10 \ \text{m}^3$~\cite{DMRadioGUT}.

The signal power is given by
\beq
\label{eq:Psig}
P_\text{sig.} \simeq  Q_\text{LC} \, \w \, \int_{V_\text{det.}} \hspace{-0.2cm} d^3 \tilde{\xv} ~ |\Ev_\text{sig.}|^2
~,
\eeq
where $E_\text{sig.}$ is the electric field sourced by $\rho_\pm$ inside the detection region and the integral is performed over the detector volume $V_\text{det.}$. We evaluate $P_\text{sig.}$ numerically for the charge densities $\rho_\pm$  discussed in Secs.~\ref{sec:chargedensity}$-$\ref{sec:perturbcharge}. We assume that the detector shield is grounded; in this case, image charges need to be included when evaluating $|\Ev_\text{sig.}|$, which we evaluate in the quasi-static limit. 

We parametrize the integral of \Eq{eq:Psig} as
\beq
\int_{V_\text{det.}} \hspace{-0.2cm} d^3 \tilde{\xv} ~ |\Ev_\text{sig.}|^2 
\equiv \frac{1}{15} ~ \eta ~ \big( R_\text{def.} ~ \rho_\pm^{(\text{Debye})}\big)^2 \, V_\text{def.}
~,
\eeq
which defines the dimensionless constant $\eta$. For a charge distribution that is spatially uniform inside the detector shield with magnitude $\rho_\pm^{(\text{Debye})}$ (see \Eq{eq:rhoDebye}), then Gauss's law gives $\eta = 1$. More generally, $\eta \neq 1$ accounts for the fact that  the MCP charge density $\rho_\pm (\tilde{\xv})$ is not spatially uniform throughout the detector volume. The precise value of $\eta$ depends on the velocity distribution of the MCPs. Evaluating $\Ev_\text{sig.}$ numerically, we find that $\eta \sim \order(10^{-2}) - \order(10)$ where the lower and upper part of this range corresponds to a maximally and minimally perturbed velocity distribution, respectively. In our sensitivity projections, we accordingly modify the particular value of $\eta$ depending on the form of the velocity distribution that is assumed. Note that although this range of $\eta$ spans three orders of magnitude, $P_\text{sig.} \propto \eta \, q_\x^4 \, n^2 \propto \eta \, q_\x^8$ implies that this only amounts to a factor of $\sim 2$ variation in the sensitivity to $q_\x$.

Stray electromagnetic fields can be efficiently attenuated by the detector shield. In our projections, we assume that the experimental reach is limited by thermal Johnson-Nyquist noise. This is the case for existing light-shining-through-wall experiments, as in, e.g., Ref.~\cite{DarkSRF}, and is expected to dominate over other forms of noise, such as fluctuations intrinsic to the readout amplifier, in future LC circuit setups such as DM Radio~\cite{Silva-Feaver:2016qhh,Godfrey:2021tvs,DMRadioGUT}. The signal-to-noise ratio is given by $\text{SNR} = P_\text{sig.} / P_\text{noise}$, where $P_\text{noise}$ is the noise power arising from thermal fluctuations. The phase and frequency of the oscillating electric field signal is determined from the deflector and the Earth velocity $v_\oplus$. Hence, if the deflector phase and frequency are monitored throughout the experimental run, this allows for a measurement of the signal amplitude, as opposed to power. If the signal phase can be determined in this manner, then $P_\text{noise} \simeq T_\text{LC} / t_\text{int.}$ in the SNR, where $T_\text{LC}$ is the temperature of the LC circuit and $t_\text{int.}$ is the total integration time of the experiment~\cite{Graham:2014sha}. In our sensitivity projections, we will assume that this is the case. We note however that if instead the deflector phase is not measured, then $P_\text{noise} \sim T_\text{LC} \sqrt{\w / (Q_\text{LC} \, t_\text{int.})}\, $; for the experimental parameters adopted here, e.g., $\w \sim 10 \ \text{kHz}$, $t_\text{int.} \sim 1 \ \text{yr}$, and $Q_\text{LC} \sim 10^7$, this corresponds to a relative increase in noise power of $\sim 200$, translating to a factor of $\sim 2$ degradation in sensitivity to millicharge $q_\x$.

The estimated sensitivity of a direct deflection setup, corresponding to $\text{SNR}> 1$, is shown in \Fig{fig:reach} compared to existing constraints (shaded gray) for fermionic MCPs and various assumptions concerning the evolution of the  solar basin (solid and dashed lines). In each case, we assume that $V_\text{def.} = V_\text{det.} = 10 \ \text{m}^3$, $\w = 10 \ \text{kHz}$, $E_\text{def.} \equiv Q_\text{def.} / (4 \pi R_\text{def.}^2) = 10 \ \text{kV} / \cm$, $Q_\text{LC} = 10^7$, $T_\text{LC} = 10 \ \text{mK}$, and $t_\text{int.} = 1 \ \text{yr}$. In constructing the solid lines and dashed lines, we assume that phase space mixing from gravitational perturbations occurs on timescales shorter or longer than $t_\odot$, respectively. 

As mentioned in \Sec{sec:chargedensity}, the deflection experimental approach relies on the MCPs being approximated as a continuum. In \Fig{fig:reach}, we enforce that $n \gtrsim 0.1 \ \cm^{-3} \times (10 \ \text{m}^3 / V_\text{def.})$, such that Poisson fluctuations in the relative number of MCPs in the deflector/detector are less than $\sim 10^{-3}$. For a MCP solar basin that saturates the upper limit from Fermi statistics in \Sec{sec:saturation}, this imposes a lower bound on the MCP mass, corresponding to the leftmost part of the sensitivity contours for each case shown in \Fig{fig:reach}. 

From the discussion above, we can understand the differences between the two sensitivity projections shown in \Fig{fig:reach}. As illustrated in \Fig{fig:xydeflect} and discussed further in \Sec{sec:perturbcharge}, the signal electric field from a MCP population that is significantly perturbed by gravitational encounters is suppressed compared to an unperturbed population, assuming comparable basin densities. As a result, a direct deflection setup has a reduced reach to a perturbed basin (compared to one that is unperturbed) for MCP masses near the upper part of the mass range shown in \Fig{fig:reach}. However, for smaller masses, a direct deflection setup is sensitive to couplings $q_\x$ such that Fermi statistics suppresses the MCP solar production rate at times before $t_\odot$. As discussed in \Sec{sec:saturation} and evident in \Fig{fig:production}, the saturation density $n_\text{sat.}$ is larger for a basin phase space that is significantly perturbed by gravitational interactions. Hence, for a perturbed basin, the deflection setup has an enhanced sensitivity to small MCP masses, significantly extending the range of masses that satisfy the continuum-criterion of the previous paragraph, $n \gtrsim 0.1 \ \cm^{-3}$, by nearly an order of magnitude.

As discussed above, the sensitivity of direct deflection to a solar basin of MCPs depends on the timescale of gravitational interactions compared to the age of the solar system. Under various assumptions, the projected sensitivity can explore viable couplings roughly an order of magnitude smaller than existing constraints for masses ranging from $\order(100) \ \text{meV}$ to $\order(100) \ \eV$. The most stringent existing constraints in this mass range are shown in gray in \Fig{fig:reach}. In dark gray, we show the limit derived from the standard solar model; a sizeable relativistic solar flux of MCPs would unacceptably modify the observed helioseismology and solar neutrinos~\cite{Vinyoles:2015khy}. Limits derived from energy loss from horizontal branch and red giant stars are shown in light gray. These are comparable to those derived from solar observations, but are sensitive to larger masses~\cite{Davidson:2000hf}. 

\section{Post-Production Interactions}
\label{sec:interactions}

In our analysis above, we assumed that once produced in the solar interior, MCPs do not significantly interact to the present day. However, for strong enough interactions with themselves or normal matter, this is no longer the case. In the remaining sections, we show that the validity of this assumption depends on the details of the particle physics model. Self-interactions, for instance, may or may not significantly modify the basin density. Regardless, as we show below, over a large region of parameter space, such dynamics have little effect on the claims of the previous sections. 

\subsection{Interactions with the Standard Model}

For strong enough interactions with the solar environment, MCPs may become trapped, or at least significantly perturbed, in their journey through the solar interior. In this section, we evaluate the impact of such processes. For instance, MCPs may Coulomb scatter off of the solar plasma. In the non-relativistic limit, the transfer cross section for fermionic MCPs scattering off of electrons is given by~\cite{Dvorkin:2019zdi}
\beq
\sigma_T \simeq \frac{16 \pi \alpha_\text{em}^2 \, q_\x^2 \, m_\x^2}{m_{D \odot}^4}
~,
\eeq
where $m_{D \odot} \simeq 6 \ \keV$ is the photon's Debye mass in the solar core. Note that $m_{D \odot}$ is much greater than the typical momentum transfer involved in the scattering, $m_\x \, v_e$, where $v_e \sim \text{few} \times 10^{-2}$ is the characteristic solar electron velocity. The MCP mean free path is then $\lambda_\text{mfp} \sim 1 / ( n_e \, \sigma_T)$, where $n_e \sim 10^{26} \ \cm^{-3}$ is the electron density in the solar core. In the parameter space of interest, the mean free path is larger than the solar radius by many orders of magnitude,
\beq
\lambda_\text{mfp} \sim \order(10^{14}) ~ r_\odot \times \bigg(\frac{q_\x}{10^{-14}} \bigg)^{-2} \bigg(\frac{m_\x}{100 \ \eV} \bigg)^{-2}
~,
\eeq
such that MCPs do not scatter in the solar interior over the entire age of the solar system.

MCPs can also couple to the solar magnetic field, $B_\odot \sim 1 \ \text{G}$. However, its influence on the MCP trajectories is model-dependent. For instance, if MCPs can be thought of as coupling directly to SM electromagnetism, their gyroradius in the solar vicinity is
\beq
r_g \sim \frac{m_\x \, v^\p}{e q_\x \, B_\odot}
~.
\eeq
Such motion does not significantly perturb the MCP trajectories if $r_g$ is larger than the coherence length of the magnetic field. If we take that length to be its maximum possible size, i.e., the solar radius $r_\odot \sim 10^6 \ \km$, this criterion is satisfied for
\beq
\label{eq:gyro1}
q_\x \lesssim \order (10^{-14}) \times \Big( \frac{m_\x}{100 \ \eV} \Big)
~,
\eeq
where we have taken the MCP velocity at production $v^\p \sim 10^{-3}$ to be comparable to the escape velocity near the Sun. If the solar magnetic field was perfectly coherent over $r_\odot$ (which is an unphysical and overly conservative assumption), then MCPs with charges above the value in \Eq{eq:gyro1} would be efficiently confined to the solar interior, as their trajectories would simply follow the solar magnetic field lines. However, the dynamics and structure of the solar interior's magnetic field is far from understood at a detailed level~\cite{kosovichev_2008}. Hence, it is not unreasonable to consider the possibility that an $\order(1)$ fraction of MCPs are able to escape the solar interior even for gyroradii much smaller than the solar radius.

Alternatively, if MCPs do not directly couple to the SM photon, but instead couple to SM currents through a light kinetically-mixed dark photon as outlined in \Sec{sec:summary}, then they only effectively couple to normal electromagnetic fields on distance scales smaller than the Compton wavelength of the dark photon; on greater distance scales, the interactions are exponentially screened. As a result, MCPs only couple to the solar magnetic field sourced within the $\Ap$ Compton wavelength. Approximating the electromagnetic currents responsible for the solar magnetic fields  as spatially uniform (which is unphysical and maximally conservative), the effective magnetic field that such MCPs couple to is suppressed by $\mAp^{-1} / r_\odot$ for dark photons shorter-ranged than the solar radius ($\mAp \gg r_\odot^{-1} \sim 10^{-15} \ \eV$)~\cite{Lasenby:2020rlf}. In this scenario, instead of \Eq{eq:gyro1}, we find that the solar magnetic field does not perturb the MCP trajectory if 
\beq
\label{eq:gyro2}
q_\x \lesssim \order(10^{-14}) \times \Big( \frac{\mAp}{10^{-15} \ \eV} \Big) \, \Big( \frac{m_\x}{100 \ \eV} \Big)
~.
\eeq
Hence, in the parameter space of interest, MCPs of mass $m_\x \sim 100 \ \text{meV}$ or $\sim 100 \ \eV$ are not perturbed by the solar magnetic field for $\mAp \gtrsim 10^{-12} \ \eV$ or $\mAp \gtrsim 10^{-15} \ \eV$, respectively. Note that this is consistent with the previous requirement that MCP interactions are long-ranged on the scale of terrestrial experiments, which requires $\mAp \lesssim 10^{-8} \ \eV$ (see \Sec{sec:summary}). Throughout this study, we assume that the dark photon is long-ranged compared to a detector on Earth yet significantly massive to screen the effect of the Sun's magnetic field, leaving a more detailed investigation to future work.

Although not the focus of this section, it is worthwhile to estimate the effects of Earth's terrestrial magnetic and electric fields on the local density of the MCP basin, since the calculation is nearly identical.\footnote{In our analysis, we ignore the effect of Earth's gravitational field, since near Earth the total gravitational potential is dominated by the solar component.} The strength of Earth's magnetic field is comparable to the Sun's. However, Earth's radius is smaller by a factor of $\sim 10^{-2}$ and the characteristic velocity of the basin near Earth is smaller by a factor of $10^{-1}$. Hence, for long-ranged MCP interactions, the requirement that Earth's magnetic field does not perturb the local density of the MCP solar basin is weaker by an order of magnitude in $q_\x$ compared to the previous upper bounds. 

Unlike the Sun, the Earth's atmosphere is an efficient insulator, thereby allowing a large potential difference of $\sim 1 \ \text{MV}$ between the ground and ionosphere on Earth, separated by $\sim 50 \ \km \sim (10^{-11} \ \eV)^{-1}$. This acts as either a potential barrier or as a well to the local MCP basin, depending on the sign of the millicharge. Approximating the terrestrial electric charge configuration as a spherical capacitor, the effect of Earth's atmospheric voltage does not effect the local MCP density provided that $e q_\x \, e^{- \mAp /  (10^{-11} \ \eV)} \times 1\,\text{MV} \lesssim m_\x \, v^2$, i.e., 
\beq
q_\x \lesssim \text{few} \times 10^{-13} \times e^{\mAp  /  (10^{-11} \ \eV)} \, \Big( \frac{m_\x}{100 \ \eV} \Big)
~,
\eeq
where we have set the MCP velocity at Earth $v \sim 10^{-4}$ to be comparable to the local solar escape velocity. Hence, we find that in a small fraction of the low-mass parameter space shown in \Fig{fig:reach}, for long-ranged interactions, the atmospheric voltage can significantly accelerate and decelerate MCPs of opposite charge, leading to a charge separation below sea-level. We note that for MCPs that are truly charged under electromagnetism, terrestrial charge separation of MCPs leads to an enhancement in the local MCP density underground. In this case, the larger underground MCP density means that a direct deflection setup placed beneath the Earth's surface would have increased sensitivity compared to one on the surface. Note that the direct deflection signal is not very sensitive to the net charge of the ambient MCP population (i.e., the expected signals from a charge-symmetric and a charge-asymmetric population are approximately the same up to factors of order unity). If, on the other hand, such interactions are mediated by a light $\Ap$, terrestrial charge separation does not continue to today, since the influx of MCPs quickly shields the dark electric field sourced by the Earth (note that the density required for this is not particularly high because MCP interactions with the atmospheric voltage are suppressed by the kinetic mixing parameter compared to MCP interactions with the population of shielding-MCPs). For these reasons, we can ignore the effect of Earth's atmospheric voltage on the direct deflection signal considered in this work. 

\subsection{Self-Annihilation}

Our discussion so far has focused mainly on processes that contribute positively to the total basin density, specifically the solar production rate in Secs.~\ref{sec:basin} and \ref{sec:solarproduction}. However, hidden sector processes can deplete this MCP population over timescales comparable to the age of the solar system, $t_\odot \simeq 4.5 \times 10^9 \ \text{yr}$, such as annihilations of MCPs into dark photons $\Ap$ or SM photons $\g$. Note that for MCP interactions mediated by a massive kinetically-mixed dark photon, MCPs solely couple to the $\Ap$. While MCPs additionally annihilate to the SM photon when the dark photon is massless, annihilations to $\Ap$ final states typically dominate since $e q_\x \ll e^\p$ for $\eps \ll 1$, leading to a relative enhancement of this channel of order $(e^\p/e q_\x)^4 \sim 1 / \eps^4$. For MCP interactions generated by means other than kinetic mixing~\cite{Batell:2005wa}, annihilations directly to photon pairs may be the leading process, but regardless are suppressed by a factor of $(e q_\x)^4$, rendering them negligible in the parameter space of interest. Hence, in this section, we solely focus on the potential implications of annihilations to an ultralight kinetically-mixed dark photon.

The cross section for MCP annihilations to dark photons is approximately
\beq
\label{eq:sigmaAnn}
\sigma_\text{ann.} v_\text{rel.} \simeq  \pi \alpha^{\p \, 2} / m_\x^2
~,
\eeq
where $v_\text{rel.}$ is the relative velocity of the MCP pair. Bound state formation via $\Ap$ emission is also possible if the dark photon is sufficiently long-ranged, but the corresponding rate is suppressed compared to that of perturbative annihilations by multiple powers of $\alpha^\p / v_\text{esc.}$~\cite{Cirelli:2016rnw}, which is much smaller than unity in the parameter space of interest.

Self-annihilations act as a sink for MCPs, and one could naively determine whether the sink is relevant for depleting the local density by comparing the rate for annihilation at the location of Earth, $n(r_\oplus) \sigma_\text{ann.} v_\text{rel.}$, to the rate of replenishment of MCPs, $\dot{n} (r_\oplus)/n(r_\oplus)$ $\sim 1 / t$ for a basin that has not yet saturated (see \Sec{sec:saturation}). However, this estimate would be neglecting the fact that along their orbits, MCPs see dramatic variations in the basin density that must be accounted for. Let us assume that the MCP solar basin is unaffected by annihilations and self-consistently show that under this assumption there exists a region of parameter space where each particle experiences fewer than one annihilation, on average, over the solar lifetime $t_\odot$.

For a given radial orbit $r(t,E, \ell)$, we integrate the interaction rate of a particle along the radial trajectory throughout the history of the solar system, such that the expected number of annihilations experienced by such a test particle of energy $E$ and angular momentum $\ell$ is
\begin{align}
\label{eq:NannihIntegral}
N_\text{ann.}(E, \ell) &\simeq \sigma_\text{ann.} v_\text{rel.} \, \int_0^{t_\odot} dt ~ t ~ \dot{n} [r(t, E, \ell)] 
\nl
&\simeq \sigma_\text{ann.} v_\text{rel.} \, \big\langle \dot{n} [r(t, E, \ell)] \big\rangle_{t_\text{orbit}} ~ \frac{t_\odot^2}{2}
~,
\end{align}
where the angle brackets denote a time-average over a single orbit and we conservatively integrate over the entire lifetime of the Sun, even though many particles in the basin have been present for only a fraction of that lifetime. In the first equality of \Eq{eq:NannihIntegral} we have approximated $n (t) \simeq \dot{n} \, t$ (self-consistently assuming that the presence of the basin does not affect the production rate) and in the second equality we have made use of the large separation of timescales to average over the short orbit timescale $t_\text{orbit} \sim \text{month}$ (for orbits with aphelia near the Earth) before integrating over the long solar timescale $t_\odot \sim \text{Gyr}$. 
In other words, we approximate the MCP density as being constant during any individual orbit, and since $\dot{n}$ at any given location is independent of time we simply average the appropriate density along the orbit and separate that from the slow filling of the solar basin.\footnote{More formally, for a function $f(t)$ evolving periodically over a characteristic timescale $\tau$ that is much less than some long integration time $T$ such that $T / \tau$ is some large integer, $\int_0^T dt \, f(t) \, t \simeq \int_0^{\tau} dt \, f(t) \, (\tau / 2) + \int_{\tau}^{2 \tau} dt \, f(t) \, (3 \tau / 2) + \cdots + \int_{T - \tau}^{T} dt \, f(t) \, \big( (2 T - \tau) / 2 \big) = (\tau / 2) \big( \int_0^\tau f(t) \big) \, \sum_{n=0}^{T/\tau - 1} (2n+1) = (T^2 / 2) ~ \langle f(t) \rangle_\tau$.}

Note that we cannot approximate the solar potential as $\Phi (r) \propto 1/r$ for portions of the orbit inside the Sun since $\Phi(r)$ is approximately independent of $r$ for $r \lesssim 0.1 \, r_\odot$ (see the rightmost panel of \Fig{fig:Sunstuff}). Therefore, we cannot assume the scaling $\dot{n} \propto r^{-4}$ of \Eq{eq:napprox1} (assuming this would result in $N_\text{ann.}$ being highly sensitive to the choice of the minimum cutoff radius when regulating the integral of \Eq{eq:NannihIntegral}). Instead, we adopt the full form for $\Phi(r)$ when evaluating $\dot{n}$ via \Eq{eq:nexact1} and when numerically solving for the orbit $r(t,E, \ell)$ (assuming no dissipation of energy). These are then used to evaluate the time averaged quantity $\langle \dot{n} [r(t, E, \ell)] \rangle_{t_\text{orbit}}$ in \Eq{eq:NannihIntegral}. Numerically, we find that for orbits with aphelia near Earth such that $E \simeq m_\x \, \Phi(r_\oplus)$, a good approximation is $\langle \dot{n} [r(t, E)] \rangle_{t_\text{orbit}} \sim 10^7 \times \dot{n} (r_\oplus)$. Using this in the result above, we then find 
\beq
\label{eq:Nannapprox}
N_\text{ann.} \sim 10^7 \times n(r_\oplus) \, \sigma_\text{ann.} v_\text{rel.} ~ t_\odot
~.
\eeq
Annihilations do not modify the MCP density at $r_\oplus$ if $N_\text{annih.} \lesssim 1$.

We note that \Eq{eq:Nannapprox} is most likely extremely conservative in that it overestimates the likelihood for a $\Ap$-coupled MCP to annihilate over a solar lifetime. In particular, we have made various simplifying assumptions that maximize the predicted self-interaction rate. For instance, we assumed that: the MCP has been gravitationally bound over the entire age of the solar system, its orbit is purely radial, the aphelia of its orbit is near Earth, and that the basin density is unsaturated by absorption and is not affected by Pauli blocking, i.e., that the density scales as $n \propto 1/r^4$ instead of $n \propto 1/r^{5/2}$ or $n \propto 1/r^{3/2}$ (see \Sec{sec:saturation}). We have also neglected additional dynamics, such as the possibility that particles on orbits with smaller aphelia annihilate before orbits with larger aphelia, thus depleting the density of target-scatterers at small radii without directly modifying the density near Earth. Each of these assumptions strengthens the likelihood for self-interactions to occur. For instance, MCPs on orbits with smaller eccentricity (due to non-zero angular momentum at production or late-time gravitational interactions) or larger aphelia, spend more time in the less dense environment at larger radii, thus softening the self-interaction rate. For purely circular orbits, the numerical prefactor in \Eq{eq:Nannapprox} (and below in \Eq{eq:Nscattapprox}) should be set to unity, such that the predicted interaction rate is reduced by several orders of magnitude. 

By substituting \Eq{eq:sigmaAnn} into \Eq{eq:Nannapprox} and taking $N_\text{ann.} \lesssim 1$, we place a conservative mass-dependent upper bound on the MCP self-coupling $\alpha^\p$. In doing so, we conservatively adopt the largest MCP densities considered in this work, which from \Fig{fig:production} corresponds to $n (r_\oplus) \sim 10^5 \ \cm^{-3}$ for $m_\x \sim 50 \ \eV$ and $q_\x \sim 10^{-14}$. In this case, we find that $\alpha^\p \lesssim 10^{-14}$. For smaller masses, the annihilation cross section grows as $\sigma_\text{ann.} v_\text{rel.} \propto 1/m_\x^2$, while the density $n \propto \mathcal{Q}_v / m_\x$ falls faster than $m_\x^2$ for fermionic MCPs (see \Sec{sec:solarproduction}). Hence, this upper bound on $\alpha^\p$ is significantly less restrictive for fermions much lighter than $\sim 50 \ \eV$.

\subsection{Self-Scattering}
\label{sec:scattering}

In addition to facilitating the annihilation processes discussed in the previous subsection, a light dark photon also mediates self-scattering of MCPs. In the weakly-coupled/Coulomb ($\alpha^\p \, \mAp \ll m_\x \, v_\text{rel.}^2$), classical ($m_\x \, v_\text{rel.} \gg \mAp$), and perturbative/Born ($\mAp \gg \alpha^\p \, m_\x$) regimes, the limiting form for the viscosity cross section\footnote{As discussed in Ref.~\cite{Colquhoun:2020adl}, the viscosity cross section is the relevant quantity for heat-conductivity, is well-defined for the scattering of either non-identical or identical MCPs, and preferentially weights scattering that significantly modifies orbital trajectories.} of MCP elastic scattering is~\cite{Knapen:2017xzo,Dvorkin:2019zdi,Colquhoun:2020adl}
\beq
\label{eq:sigmascatt}
\sigma_V \, v_\text{rel.} \to \frac{32 \pi \, \alpha^{\p \, 2}}{m_\x^2 \, v_\text{rel.}^3} ~ \log{\left( \frac{m_\x \, v_\text{rel.}}{\mAp} \right)}
~.
\eeq
As can be seen by comparing \Eq{eq:sigmascatt} to \Eq{eq:sigmaAnn} in the previous subsection, MCP self-scattering is parametrically enhanced in the low velocity limit by $\sim v_\text{rel.}^{-3} \gg 1$ compared to annihilations. The estimate of the number of scattering events experienced by a MCP on a radial orbit is similar to that in \Eq{eq:NannihIntegral} in the previous subsection. However, the enhancement of the scattering rate in larger density environments at smaller radii is tempered by the corresponding larger velocities, which suppresses the scattering cross section. 

Following the procedure outlined in Ref.~\cite{Colquhoun:2020adl}, the rate at which MCP self-scattering leads to significant energy transfer between a test particle with velocity $v_\text{traj.}$ with respect to the Sun (note this is distinct from $v_\text{rel.}$, the relative velocity between MCPs in the basin) and the rest of the basin population is
\beq
\label{eq:Gammascatt}
\Gamma_\text{scatt.} \simeq \frac{n}{4 v_\text{traj.}^2} \, \langle \sigma_V \, v_\text{rel.}^3 \rangle_\text{basin}
~,
\eeq
where the angle brackets denote an average over the basin phase space. We conservatively adopt a basin phase space distribution that is unperturbed by gravitational encounters, as in \Sec{sec:basin}. In evaluating the basin-average of \Eq{eq:Gammascatt}, the approximate form for $\sigma_V$ in \Eq{eq:sigmascatt} is not valid over the entire velocity range. Instead, we use the complete set of semi-analytic expressions of Ref.~\cite{Colquhoun:2020adl}, which is especially important in regulating the rate at small velocities. 

Similar to \Eq{eq:NannihIntegral}, the scattering rate $\Gamma_\text{scatt.} (r)$ allows us to determine the number of scatters experienced by a test particle of energy $E$ and angular momentum $\ell$ along an orbit $r(t,E, \ell)$ with velocity $v_\text{traj.} (t, E, \ell)$, 
\begin{align}
\label{eq:NscattIntegral}
& N_\text{scatt.}(E, \ell) \simeq \int_0^{t_\odot} dt ~ \Gamma_\text{scatt.} [r(t,E)] 
\nl
&\simeq \bigg\langle \frac{\dot{n} [r(t, E, \ell)]}{v_\text{traj.} (t, E, \ell)^2} ~ \langle \sigma_V \, v_\text{rel.}^3 \rangle_\text{basin} \bigg\rangle_{t_\text{orbit}} ~ \frac{t_\odot^2}{8}
~.
\end{align}
Numerically, we find that for most orbits with aphelia near Earth \Eq{eq:NscattIntegral} is well approximated by 
\beq
\label{eq:Nscattapprox}
N_\text{scatt.} \sim 10^4 \times n (r_\oplus) \, (  \sigma_V v_\text{rel.} )_\oplus ~ t_\odot
~,
\eeq
where for concreteness we have chosen $\mAp = 10^{-8} \ \eV$ since this corresponds to the largest value of $\mAp$ that we consider in this work (we are not particularly sensitive to this choice since $\sigma_V$ only has a mild logarithmic dependence on $\mAp$), and $( \sigma_V v_\text{rel.} )_\oplus$ is defined to be the simple limiting form of the scattering cross section in \Eq{eq:sigmascatt} with $v_\text{rel.} \to v_\text{esc.} (r_\oplus)$. MCPs with aphelia near Earth do not scatter over the lifetime of the solar system if $N_\text{scatt.} \lesssim 1$. Comparing Eqs.~(\ref{eq:sigmascatt}) and (\ref{eq:Nscattapprox}) to Eqs.~(\ref{eq:sigmaAnn}) and (\ref{eq:Nannapprox}) of the previous subsection, the number of scatters per annihilation is $N_\text{scatt.} / N_\text{annih.} \sim \big( 32 / v_\text{esc.}(r_\oplus)^3 \big) \, \big(10^4 / 10^7 \big) \sim 10^{10}$. Hence, appropriately rescaling the bound on  $\alpha^\p$ from the previous subsection, we find that $N_\text{scatt.} \lesssim 1$ corresponds to $\alpha^\p \lesssim 10^{-18}$. As discussed previously in \Sec{sec:summary}, in theories involving light MCPs, small values of $\alpha^\p$ are in fact motivated from considerations of stellar energy loss.

\Eq{eq:Nscattapprox} and the resulting upper bound on $\alpha^\p$ are extremely conservative for the same reasons as discussed in the previous subsection for the MCP annihilation rate. However, unlike annihilations, the probability for MCP self-scattering to efficiently transfer momentum between particles is additionally suppressed by the fact that Pauli-blocking is significant for a nearly-degenerate basin phase space (see \Sec{sec:saturation}). Hence, we expect a more accurate treatment of scattering to significantly relax these upper bounds on $\alpha^\p$. 

We have determined the size of $\alpha^\p$ for which most MCPs self-scatter over a solar lifetime. As an aside, we may now ask: what happens if such scattering does occur? Although scattering does not directly alter the total number of MCPs in the solar basin, once energy is efficiently transported throughout the basin, the MCP population enters a state of hydrostatic equilibrium. An investigation of these dynamics is presented in \App{app:hydro}, where we show that the phase space properties of a basin in hydrostatic equilibrium are significantly modified, potentially resulting in a  reduction of the local MCP density. 

\section{Discussion and Conclusions}
\label{sec:conclusion}

We have outlined a new approach to detect millicharged particles emitted from the Sun. In a small part of phase space, such particles are produced with sufficiently small velocities to remain gravitationally bound to the solar system over billions of years, constituting a ``solar basin"~\cite{VanTilburg:2020jvl,Lasenby:2020goo}. Traditional direct detection techniques, such as searches for elastic scattering that deposits more than $ 1 \ \text{eV}$ of kinetic energy onto a target, are incapable of detecting such a population due to the small mass and velocities of millicharged particles in the solar basin. 

We have shown that a helioscope consisting of a ``direct deflection" setup is a promising avenue to overcome these difficulties. Applied to this scenario, the experimental approach involves inducing collective disturbances into the background of millicharged particles which can be resonantly detected with precision sensors~\cite{Berlin:2019uco}. Crucially, this setup lacks a classical kinematic threshold, since the ability to induce collective effects into the basin is parametrically enhanced by the small velocity of gravitationally bound particles. Our study indicates that a resonant detector consisting of a $\sim$meter-sized cryogenic LC circuit, similar to the one being developed for the experiment DM Radio~\cite{Silva-Feaver:2016qhh, Godfrey:2021tvs,DMRadioGUT}, holds promising sensitivity to millicharged particles in the $\text{eV}-\text{keV}$ mass range and with couplings well below existing lab-based or astrophysical constraints. This same setup can operate concurrently as a search for sub-GeV dark matter~\cite{Berlin:2019uco}; a setup optimized for detecting a millicharged solar basin and pointing along the direction of the basin wind would still be sensitive to millicharged dark matter. 

In this work, we have for the first time determined the phase space density of the basin that is imparted onto it from production processes in the Sun. This is needed for a detailed understanding of the experimental signal discussed here and more generally is useful for a complete understanding of stellar basins. For instance, a detailed understanding of the phase space has allowed us to identify new dynamics associated with millicharge solar basins (and related models in which solar production requires at least two dark sector particles in the final state). In this case, unlike solar basins consisting of, e.g., axions or dark photons~\cite{VanTilburg:2020jvl,Lasenby:2020goo}, the role of solar absorption is greatly diminished. As a result, the rate for pair-production of light bosons may continue to grow exponentially until the phase space occupancy of the solar basin is highly degenerate. We leave a more detailed investigation of this effect and its potential consequences to future work~\cite{KVT}.

Although we have focused solely on solar production in this study, it may also be worth pursuing the detection of the terrestrial basin of millicharged particles that are produced in Earth's core and remain bound through Earth's gravitational and/or electromagnetic fields. Millicharge basins surrounding other stellar systems, such as neutron stars and white dwarfs, may also have interesting implications since their densities are enhanced by the steeper gravitational potential and strong trapping ability of astrophysical electromagnetic fields.

\begin{acknowledgments}
We thank Saniya Heeba, Cristina Mondino, Ken Van Tilburg, Aaron Vincent, and  Hai-Bo Yu for valuable discussions. AB was supported by a James Arthur Fellowship. KS was supported by a Natural Sciences and Engineering Research Council of Canada
(NSERC) Subatomic Physics Discovery Grant, by a Pappalardo Fellowship in the MIT Department of Physics, and by NASA through the NASA Hubble Fellowship grant HST-HF2-51470.001-A awarded by the Space Telescope Science Institute, which is operated by the Association of Universities for Research in Astronomy, Incorporated, under NASA contract NAS5-26555.
\end{acknowledgments}

\bibliographystyle{utphys}
\bibliography{main}

\onecolumngrid
\newpage
\appendix

\section{Solar Production of Millicharged Particles}
\label{app:MCPproduction}
In this appendix, we derive the plasmon decay rate function $\mathcal{Q}_v (\gamma_{T, L}^*)$, which was previously discussed in \Sec{sec:solarproduction}. The MCP luminosity per unit solar volume $Q$ is related to the matrix element $\M$ for plasmon decay by
\beq
\label{eq:Qgen}
Q = 2 \, \int d \Pi_{\g^*} \, d\Pi_\x \, d\Pi_{\xb} ~ \overline{|\M|^2} ~ (2 \pi)^4 \delta^4 (p_\x + p_{\xb} - p_{\g^*}) ~ \w_\x ~ \mathcal{F}
~,
\eeq
where we have included a factor of two to account for the sum of both $\text{MCP}$ and $\overline{\text{MCP}}$ populations (since MCPs are produced symmetrically, the energy loss rate is the same for antiparticles, denoted symbolically with an overline/bar). The four-momentum of species $i = \g^*, \text{MCP}, \overline{\text{MCP}}$ is denoted as $p_i^\mu = (\w_i, \kvec_i)$ and $\overline{|\M|^2}$ is the  squared matrix element for $\g^* \to \text{MCP} ~ \overline{\text{MCP}}$ averaged over the initial and final spins. Above, the Lorentz invariant phase space elements are defined as
\beq
d\Pi_i \equiv \frac{g_i}{(2 \pi)^3} \, \frac{d^3 \kvec_i}{2 \w_i}
~,
\eeq
where $g_i$ are the number of internal degrees of freedom. The phase space distribution functions $f_i = f_i (\w_i)$ are incorporated in \Eq{eq:Qgen} as
\beq
\label{eq:F}
\mathcal{F} \equiv f_{\g^*} \, (1 \pm f_\x)\, (1 \pm f_{\xb}) - f_\x \, f_{\xb} \, (1+f_{\g^*})
~,
\eeq
where $\pm$ corresponds to bosonic/fermionic MCPs, respectively. We will assume that the phase space density of MCPs is initially small, i.e., $f_\x , f_{\xb} \ll 1$, such that we can approximate $\mathcal{F} \simeq f_{\gamma^*} (\w_{\g^*}) = 1 / (e^{\w_{\g^*} / T_\odot} - 1)$.

Let us now proceed by evaluating the expression in \Eq{eq:Qgen}. Since we will be interested in the production of gravitationally bound MCPs, we work in the non-relativistic limit by taking $k_\x \simeq m_\x \, v^\p$, where $v^\p$ is the MCP velocity at the solar radius where it is produced (the other MCP of the emitted pair is produced relativistically, as implied by energy-momentum conservation). We find
\beq
\label{eq:dQdw1}
\frac{dQ}{d v^\p} \simeq \frac{g_\g \, g_\x^2}{16 \pi^3} ~m_\x^2 \, v^\p \int_0^\infty d k_{\g^*} ~ \Theta (1 - |\cos{\bar{\theta}} \, |) ~ \frac{k_{\g^*}}{\w_{\g^*}} ~ \mathcal{F} ~  \overline{|\M|^2}
~,
\eeq
where $g_\x = g_{\xb} = 2 \, (1)$ for fermion (scalar) MCPs, $\Theta$ is the Heaviside step function, 
\beq
\label{eq:costhb1}
\cos{\bar{\theta}} \equiv \frac{2 \w_{\g^*} \, \w_\x - (\w_{\g^*}^2 - k_{\g^*}^2)}{2 k_{\g^*} \, k_\x}
\eeq
is the cosine of the angle between $\kvec_\g$ and $\kvec_\x$, and $p_\x^\mu$ is the four-momentum of the non-relativistic MCP.

\Eq{eq:dQdw1} is the general expression for non-relativistic production of MCPs from plasmon decay. To further evaluate the integral over $k_{\g^*}$ in \Eq{eq:dQdw1}, we need to calculate the matrix element $\M$ as well as the dispersion relation between $\w_{\g^*}$ and $k_{\g^*}$, which depend on the spin of the MCP (spin-0 or spin-1/2) and the polarization of the plasmon (transverse or longitudinal). Here, we explicitly evaluate \Eq{eq:dQdw1} for each of these cases.

Let us first consider the decay of \emph{transverse} plasmons to MCPs. In a non-relativistic plasma such as the Sun, the transverse plasmon dispersion relation is approximately $\w_{\g^*}^2 \simeq k_{\g^*}^2 + \w_p^2$, where the plasma mass squared is $\w_p^2 = 4 \pi \alpha n_e / m_e +\mathcal{O}(T/m_e)$. With this dispersion relation, \Eq{eq:dQdw1} simplifies to
\beq
\label{eq:dQdw1b}
\frac{dQ (\g_T^*)}{d v^\p} \simeq \frac{g_\x^2}{8 \pi^3} ~m_\x^2 \, v^\p \int_{\w_p}^\infty d \w_{\g^*} ~ \Theta (1 - |\cos{\bar{\theta}} \, |) ~ \mathcal{F} ~  \overline{|\M|^2}
~.
\eeq
For the production of non-relativistic MCPs, the Heaviside step function in \Eq{eq:dQdw1b} only has weight in a small region of phase space. In particular, in the non-relativistic limit we find that $\w_{\g^*}$ is restricted to an interval characterized by a central value $\w_{\g^*} \simeq \w_p^2 / 2 m_\x$ and width $\Delta \w_{\g^*} \simeq 2 \, \w_{\g^*} \, v^\p \, (1 - 4 m_\x^2 / \w_p^2)^{1/2}$. Since $\Delta \w_{\g^*} \ll \w_{\g^*}$ for $v^\p \ll 1$, the integral in \Eq{eq:dQdw1b} can be approximated analytically, yielding
\beq
\label{eq:QvT1}
\mathcal{Q}_v (\g_T^*) \simeq \frac{g_\x^2}{32 \pi^4} ~ m_\x \, \w_p^2 ~  ( 1 - 4 m_\x^2 / \w_p^2 )^{1/2} ~ f_{\g^*} (\w_p^2 / 2 m_\x) ~  \overline{|\M|^2}
~,
\eeq
where $\mathcal{Q}_v$ is defined as in \Eq{eq:Qvdef} with index $\ind = 1$. In the non-relativistic limit, the spin-averaged squared matrix element for transverse plasmon decay to a pair of fermionic MCPs is $\overline{|\M|^2} \simeq 4 \pi \, \alpha_\text{em} \, q_\x^2 ~ \w_{\g^*} \, m_\x$~\cite{Dvorkin:2019zdi}. Using this in \Eq{eq:QvT1} gives
\beq
\label{eq:QvT2}
\mathcal{Q}_v (\g_T^* \to \text{fermionic MCPs}) \simeq \frac{\alpha_\text{em} \, q_\x^2}{4 \pi^3} \, m_\x \, \w_p^4 ~  ( 1 - 4 m_\x^2 / \w_p^2 )^{1/2} ~ f_{\g^*} (\w_p^2 / 2 m_\x)
~.
\eeq
Instead if the MCP is a scalar, we find that $\overline{|\M|^2} \propto v^{\p \, 2} \ll 1$ in the non-relativistic limit and therefore the rate is parametrically suppressed,
\beq
\mathcal{Q}_v (\g_T^* \to \text{scalar MCPs}) \propto v^{\p \, 2} 
~.
\eeq

For the decays of longitudinal plasmons, the calculation is similar to the previous one, except that the dispersion relation is $\w_{\g^*} \simeq \w_p$. In this case, the step function in \Eq{eq:dQdw1} enforces that $k_{\g^*}$ is restricted to an interval characterized by a central value $k_{\g^*} \simeq \w_p ~ (1 - 2 m_\x / \w_p)^{1/2}$ and width $\Delta k_{\g^*} \simeq 2 m_\x \, v^\p$. \Eq{eq:dQdw1} can hence be evaluated analytically such that
\beq
\label{eq:dQdw5}
\mathcal{Q}_v(\gamma_L^*) \simeq \frac{g_\x^2}{32 \pi^4} ~ m_\x^3 ~ (1 - 2 m_\x / \w_p)^{1/2} ~ f_{\g^*} (\w_p) ~  \overline{|\M|^2}
~.
\eeq
In the non-relativistic limit, the spin-averaged squared matrix element for longitudinal plasmon decay to fermionic MCPs is $\overline{|\M|^2} \simeq 4 \pi \, \alpha_\text{em} \, q_\x^2 ~ m_\x ~ (\w_{\g^*} / k_{\g^*})^2 \, (\w_{\g^*} - 2 m_\x)$~\cite{Dvorkin:2019zdi}. Using this in \Eq{eq:dQdw5} gives
\beq
\label{eq:dQdw6}
\mathcal{Q}_v(\g_L^* \to \text{fermionic MCPs}) \simeq \frac{\alpha_\text{em} \, q_\x^2}{2 \pi^3} \, m_\x^4 \, \w_p \, (1 - 2 m_\x / \w_p)^{1/2} \, f_{\g^*}(\w_p)
~.
\eeq
Instead, for scalar MCPs, the matrix element is $\overline{|\M|^2} \simeq 4 \pi \, \alpha_\text{em} \, q_\x^2 ~ (\w_{\g^*} / k_{\g^*})^2 \, (\w_{\g^*} - 2 m_\x)^2$, which yields
\beq
\label{eq:dQdw7}
\mathcal{Q}_v (\gamma_L^* \to \text{scalar MCPs}) \simeq \frac{\alpha_\text{em} \, q_\x^2}{8 \pi^3} \, m_\x^3 \, \w_p^2 \, (1 - 2 m_\x / \w_p )^{3/2} \, f_{\g^*}(\w_p)
~.
\eeq

\section{Millicharge Overdensities}
\label{app:chargedensity}

In \Sec{sec:deflection}, we discussed the MCP charge overdensities sourced by the driven electric field of the deflector. In this appendix, we now provide additional technical details for the calculations of \Sec{sec:deflection}. The general expression for the induced MCP charge overdensity $\rho_\pm$ was provided in \Eq{eq:deflectiongeneral}. For a driven deflector charge configuration consisting of a point charge surrounded by a grounded spherical shield (as described by \Eq{eq:rhodefpoint}), \Eq{eq:deflectiongeneral} can be rewritten as
\beq
\label{eq:rhopointandshell}
\rho_\pm (\tilde{\xv}, t) \simeq - \frac{(e q_\x)^2}{m_\x} ~ e^{i \w t} ~ Q_\text{def.} \, \Big( I_\text{point} (\tilde{\xv}) + I_\text{shell} (\tilde{\xv}) \Big) 
~,
\eeq
where we have defined
\begin{align}
\label{eq:rhodefbreakdown}
I_\text{point} (\tilde{\xv}) &\equiv \int d v ~  \frac{f(r_\oplus, v\, \hat{\tilde{\xv}} + \vv_\oplus)}{|\tilde{\xv}|} 
\nl
I_\text{shell} (\tilde{\xv}) &\equiv - \frac{1}{4 \pi R_\text{def.}^2} \, \int d v ~ \int_{V_\text{def.}} \hspace{-0.2cm} d^3 \tilde{\xv}^\p ~ f(r_\oplus, v\, \hat{\vv} + \vv_\oplus) ~ \frac{\delta (|\tilde{\xv}^\p| - R_\text{def.}) }{|\tilde{\xv} - \tilde{\xv}^\p|}
~.
\end{align}
The integrals $I_\text{point}$ and $I_\text{shell}$ correspond to the point charge and spherical shell contributions of the deflector charge configuration of \Eq{eq:rhodefpoint}.

At distances far from the deflector region ($\tilde{r} \gg R_\text{def.}$), the expression in \Eq{eq:deflectiongeneral} is analytically tractable. As shown in Ref.~\cite{Berlin:2019uco}, in this far-field limit $\rho_\pm \propto \mathcal{R}_\text{def.}^2$, where $\mathcal{R}_\text{def.}^2 = - Q_\text{def.} \, R_\text{def.}^2$ is the charge radius squared of the deflector, such that
\beq
\rho_\pm (\tilde{\xv}, t) \xrightarrow{\tilde{r} \gg R_\text{def.}} - \, \frac{(e q_\x)^2}{6 m_\x} ~  \mathcal{R}_\text{def.}^2 ~e^{i \w t} \int d v ~ \grad^2 \bigg( \frac{f(r_\oplus, v\, \hat{\tilde{\xv}} + \vv_\oplus)}{\tilde{r}} \bigg) 
~.
\eeq
For an anisotropic velocity distribution given by a Maxwellian shifted by the $\vv_\oplus$ ``wind," the integral over velocity in the expression above is parametrically of size $\sim - 1/(\tilde{r}^3 v_\oplus^2)$, leading to
\beq
\rho_\pm (\tilde{\xv}) \sim -  m_{D, \x}^2 \, (Q_\text{def.} / 4 \pi R_\text{def.} ) \, ( R_\text{def.} / \tilde{r} )^3
~. 
\eeq
The first two factors are the general expectation from normal Debye screening. The $1/\tilde{r}^3$ falloff in the last factor is the non-trivial modification arising from the fact that the deflector point charge is surrounded by a conducting shield of opposite charge, which screens the signal at large distances.

Let us now discuss the explicit evaluation of the induced MCP charge density $\rho_\pm$ for a few different basin velocity distributions. For a basin velocity distribution that is significantly perturbed by gravitational interactions, as in \Sec{sec:perturbcharge}, we adopt the Gaussian distribution of \Eq{eq:gvperturb} and evaluate \Eq{eq:rhodefbreakdown} numerically. Alternatively, for a MCP velocity distribution that is unperturbed by gravitational or hidden sector interactions, as in \Sec{sec:unperturbcharge}, we approximate the velocity distribution in the lab frame using \Eq{eq:unpertf}, i.e., $f (r_\oplus, \vv + \vv_\oplus) \simeq n (r_\oplus) \, \delta (v_x - v_\oplus) ~ \delta (v_y) ~ g_z (v_z)$, where we have defined 
\beq
\label{eq:gz}
g_z (v_z) \equiv \frac{1}{2 v_\text{esc.}(r_\oplus)} ~ \Theta(v_\text{esc.}(r_\oplus) - |v_z|)
~.
\eeq
In this case, for the contribution from the shielded point charge of the deflector, the integral over $v$ in the first line of \Eq{eq:rhodefbreakdown} can be evaluated analytically, yielding
\beq
\label{eq:Ipoint}
I_\text{point} (\tilde{\xv}) = \frac{1}{v_\oplus} ~ g_z \left(\frac{\tilde{z}}{\tilde{x}} ~ v_\oplus \right) \, \Theta (\tilde{x}) \, \delta (\tilde{y})
~.
\eeq
Since \Eq{eq:gz} implies that $g_z \sim 1 / \big( 2 v_\text{esc.} (r_\oplus) \big) \sim 1 / (2 \, v_\oplus)$, from \Eq{eq:rhopointandshell} we see that this will contribute a MCP surface charge density as shown in \Eq{eq:surfacecharge}. Next, to evaluate the contribution from the spherical shield of the deflector, the second line of \Eq{eq:rhodefbreakdown} simplifies to
\beq
I_\text{shell} (\tilde{\xv}) = - \frac{\Theta(R_\text{def.} - |\tilde{y}|)}{4 \pi R_\text{def.} \, v_\oplus} ~ \int_{- \sqrt{R_\text{def.}^2 - \tilde{y}^2}}^{\min{(\tilde{x}, \sqrt{R_\text{def.}^2 - \tilde{y}^2})}} ~ \frac{d\tilde{x}^\p}{\tilde{\mathcal{Z}}} ~ \Bigg[ \, g_z \left( \frac{\tilde{z} + \tilde{\mathcal{Z}}}{\tilde{x} - \tilde{x}^\p} ~ v_\oplus \right) + g_z \left( \frac{\tilde{z} - \tilde{\mathcal{Z}}}{\tilde{x} - \tilde{x}^\p} ~ v_\oplus \right) \, \Bigg]
~,
\eeq
where we defined $\tilde{\mathcal{Z}} \equiv \sqrt{R_\text{def.}^2 - (\tilde{x}^{\p 2} + \tilde{y}^2)} \, $. To estimate the characteristic size of $I_\text{shell} (\tilde{\xv})$, let us evaluate it at $\tilde{x} > R_\text{def.}$, $\tilde{y} = \tilde{z} = 0$, approximating $v_\text{esc.} (r_\oplus) \sim v_\oplus$. This yields $I_\text{shell} \sim - 1 / (4 \, R_\text{def.} \, v_\oplus^2)$. From \Eq{eq:rhopointandshell} we see that this will contribute a MCP charge density as shown in \Eq{eq:shellcharge}.

\section{Hydrostatic Equilibrium}
\label{app:hydro}

Although, unlike annihilations,  scattering is not a direct energy sink, the distribution of the MCP population is sensitive to such effects. If $N_\text{scatt.} \gg 1$, then the solar population of MCPs thermalizes, similar to the dynamics considered in models of self-interacting dark matter in which scattering equilibrates the innermost regions of galactic halos~\cite{Tulin:2017ara}. Upon thermalization, the basin MCP population approaches hydrostatic equilibrium, 
\beq
\label{eq:Euler}
\grad P_\x \simeq - m_\x \, n \, \grad \Phi
~.
\eeq
Approximating the MCPs as a non-relativistic ideal gas of temperature $T_\x(r)$, the pressure is $P_\x \simeq T_\x \, n \, $. Note that we have assumed that the MCP basin does not saturate Pauli-Dirac statistics, such that we can ignore additional contributions to $P_\x$ stemming from degeneracy pressure. After settling into hydrostatic equilibrium, we additionally model the MCPs as obeying the polytropic equation of state $P_\x \propto n^\g$, where $\g \simeq 5/3$ is the corresponding polytropic index for an isentropic monatomic gas. The ideal gas law combined with the polytropic equation of state implies that 
$\grad P_\x \simeq (5/2) \, n \, \grad T_\x$, which upon substituting into the Euler equation in \Eq{eq:Euler} gives
\beq
\label{eq:temp1}
T_\x \simeq - (2/5) \, m_\x \, \Phi
~.
\eeq
Note that this determines the mean MCP energy at radius $r$ to be 
$\langle E_\x (r) \rangle \simeq - (2/5) \, m_\x \, \Phi (r)$, corresponding to a typical speed that is below the gravitational escape velocity. 
Regardless, MCPs in the high velocity tail of the Maxwell-Boltzmann distribution have sufficient energy to escape, leading to partial evaporation of the solar basin. We discuss this below, but in order to first address the effects of hydrostatic equilibrium that are independent of evaporation, we first assume that the hydrostatic population is efficiently bound to the solar system. 

The temperature profile of \Eq{eq:temp1} implies that the resulting MCP number density scales as
\beq
n \propto \Phi^{3/2} \propto r^{-3/2}
~,
\eeq
for $r \gtrsim r_\odot$. Note that $n$ falls less steeply in heliocentric radius $r$ compared to the \emph{initially unequilibrated} density  $n^{(i)} \propto r^{-4}$ (this latter scaling assumes that the phase space has not yet been saturated, as discussed in \Sec{sec:saturation}). We fix the proportionality constant in $n$ by demanding that the total number of particles are unchanged before and after equilibration, i.e.,
\beq
\int d^3 \xv ~ n  = \int d^3 \xv ~ n^{(i)}
~, 
\eeq
where $n^{(i)}$ is the initial density before scattering occurs, as in Eqs.~(\ref{eq:nexact1}) and (\ref{eq:napprox1}). In order to analytically evaluate the integrals in the above expression, we take $\Phi \propto 1/r$. In this case, the integrals diverge at small radii, which then requires regulating the integrals by restricting  $r > r_\Phi$. We take this to be smallest radius at which the gravitational potential of the Sun is well approximated by that of a point mass, $r_\Phi \sim 0.1 \times r_\odot$. We also note that the integral over the hydrostatic density diverges at large radii. We therefore regulate the integral by taking $r < r_\text{hydro} \sim \min{(10^7 \ \text{AU} \, , \, r_\text{scatt.})}$; $10^7 \ \text{AU}$ is the maximum distance that a gravitationally bound particle could have traveled over the lifetime of the solar system and  $r_\text{scatt.}$ is defined to be the radius of the last scattering surface, i.e., the point at which a radially outward propagating MCP has an optical depth smaller than unity (see, e.g., Ch.~5 of Ref.~\cite{Catling:2017}),
\beq
\label{eq:opticaldepth}
\int_{r_\text{scatt.}}^{10^7 \ \text{AU}} dr ~ n (r) \, \sigma_V (r) \sim 1
~,
\eeq
where $ \sigma_V$ is the viscosity cross section (see \Sec{sec:scattering}).
In this sense, $r_\text{scatt.}$ corresponds to the point beyond which the MCP basin is no longer in hydrostatic equilibrium; for $r > r_\text{scatt.}$, the basin instead consists of free-streaming particles, which we assume makes up a negligible fraction of the total density compared to the hydrostatic population.

Following this procedure, if the initial unequilibrated density is not yet saturated ($n^{(i)} \lesssim n_\text{sat.}$ where $n_\text{sat.}$ is defined in \Sec{sec:saturation}), then $n^{(i)} \propto r^{-4}$ and the hydrostatic MCP density (normalized by the initially unequilibrated density)  at radius $r$ is
\beq
\frac{n (r)}{n^{(i)}(r)} \simeq \frac{3}{2} \, \left( \frac{r}{r_\text{scatt.}} \right)^{3/2} \left( \frac{r}{r_\Phi} \right)
~,
\eeq
where we have assumed $r_\text{scatt.} \gg r_\Phi$. 
Near Earth, the relative change to the local density is therefore
\beq
\label{eq:localhydro}
\frac{n (r_\oplus)}{n^{(i)} (r_\oplus)} \sim \left( \frac{220 \ \text{AU}}{r_\text{scatt.}} \right)^{3/2}
~.
\eeq
Therefore, for a last scattering surface $r_\text{scatt.} \gg 200 \ \text{AU}$, thermalization leads to a suppression in the local MCP density. Instead, note that if the phase space density of the basin has been gravitationally perturbed (as discussed in \Sec{sec:saturation}) before self-scattering drives the population towards hydrostatic equilibrium, then $n^{(i)} \propto r^{-3/2}$ from \Eq{eq:satpert}. Since this scaling is the same as that of the hydrostatic population,  
we see that scattering does not alter the radial profile of a gravitationally perturbed basin population. 

As shown in \Sec{sec:interactions}, there is a wide range in which $\alpha^\p$ is sufficiently large such that scattering may modify the distribution of the MCP solar basin, yet sufficiently small such that annihilations do not deplete the overall density. Note that when MCP self-scattering is classical and perturbative such that the expression for $\sigma_V$ in \Eq{eq:sigmascatt} is valid, $v_\text{rel.} \sim v_\text{esc.} (r) \sim 1/r^{1/2}$ implies that $n \, \sigma_V \propto r^{1/2}$, i.e., the hydrostatic basin is more tightly coupled at larger radii. As a result, we expect that $r_\text{scatt.} \gg 200 \ \text{AU}$ for $\alpha^\p \gg 10^{-18}$, and thus from \Eq{eq:localhydro} scattering may lead to a strong suppression of the local density ($n \ll n^{(i)}$) at Earth. 

In order to simplify the analysis above, we ignored scattering-induced evaporation of the thermalized population of MCPs. The incorporation of this effect leads to an additional suppression of the local MCP density after reaching hydrostatic equilibrium, analagous to how Jeans thermal escape in planetary atmospheres can deplete the abundance of lighter elements~\cite{Catling:2017}. This can be modeled by estimating the fraction of MCPs in a  Maxwellian distribution at temperature $T_\x$ that are outwardly traveling at a speed above the solar escape velocity at the last scattering surface $r \sim r_\text{scatt.}$, since such particles can travel into the collisionless regime of the solar basin ($r \gtrsim r_\text{scatt.}$) and free-stream out to infinity. The outgoing flux of such MCPs is approximated as $j_\text{evap.} \simeq n (r_\text{scatt.}) \, \langle v_\text{out,esc.} (r_\text{scatt.}) \rangle$, where $\langle v_\text{out,esc.} (r_\text{scatt.}) \rangle$ is the thermally-averaged radially-outward velocity of particles with sufficient energy to escape the solar system at the last scattering surface, i.e.,
\beq
\langle v_\text{out,esc.} \rangle \simeq \int_{v > v_\text{esc.}} \hspace{-0.2 cm} d^3 \vv ~ g_\text{eq.}(v) ~ v \, \cos{\theta}
= \frac{v_0 \, (1 + v_\text{esc.}^2 / v_0^2)}{2 \sqrt{\pi}} ~ e^{- v_\text{esc.}^2 / v_0^2} \Bigg|_{r = r_\text{scatt.}}
~.
\eeq
Above, $g_\text{eq.}(v)$ is the unit-normalized Maxwellian distribution with dispersion $v_0 = \sqrt{2 \, T_\x (r_\text{scatt.}) / m_\x}$ and $v_\text{esc.} = \sqrt{2 \, G \, M_\odot / r_\text{scatt.}}\, $ is the escape velocity at the last scattering surface. This outgoing flux depletes the solar basin density at a rate of
\beq
\dot{n}_\text{evap.} \simeq \frac{j_\text{evap.}  \, 4 \pi r_\text{scatt.}^2}{(4 \pi / 3) \, r_\text{scatt.}^3} 
= \frac{3 \, n}{2 \sqrt{\pi}} \,\frac{ v_0 \, (1 + v_\text{esc.}^2 / v_0^2)}{r_\text{scatt.}} ~ e^{- v_\text{esc.}^2 / v_0^2} \Bigg|_{r = r_\text{scatt.}}
~.
\eeq
Since $v_\text{esc.} / v_0$ is $\order(1)$ for all radii, the evaporation rate is only suppressed by the small value of $n (r_\text{scatt.})$ and the large value of $r_\text{scatt.}$. Hence, we expect evaporation of the hydrostatic population to be relevant. 

\end{document}